\documentclass[10pt,aps,prd,twocolumn,floatfix,nofootinbib,superscriptaddress]{revtex4-2}
\usepackage{amsmath,amssymb}
\usepackage{graphicx}
\graphicspath{{figures/}}
\usepackage[hidelinks]{hyperref}

\begin{document}

\title{Origin of flavour cancellations in thermal leptogenesis}
\author{Teruyuki Kitabayashi}
\email{teruyuki@tokai.ac.jp}
\affiliation{Department of Physics, Tokai University, 4-1-1 Kitakaname, Hiratsuka, Kanagawa 259-1292, Japan}

\begin{abstract}
 In thermal leptogenesis with three hierarchical heavy neutrinos $N_1$, $N_2$ and $N_3$, the baryon asymmetry can be the small remainder of large contributions that cancel. Interference terms in the CP asymmetry are known to contribute to the final asymmetry through flavour-dependent washout. However, it remains unclear where the cancellation occurs, how large the interference contribution is, and how the accuracy of approximate equations changes when this information is lost. We split the CP asymmetry into diagonal and off-diagonal parts in the charged-lepton basis and evolve them with Standard Model transport equations. We find that the cancellation occurs mainly between the diagonal and off-diagonal parts of the same heavy neutrino, and that it is caused by flavour-dependent washout. We also find that the interference between the lepton state produced in the decay and the states orthogonal to it typically accounts for most of the $N_2$ and $N_3$ contribution. Approximations that reduce the CP asymmetry to flavour populations turn out to miss most of this contribution, whereas density-matrix equations with the full CP-asymmetry matrix agree much better.
\end{abstract}
\maketitle

\section{Introduction}\label{sec:intro}

The baryon asymmetry of the Universe may originate from the CP-violating decays of heavy Majorana neutrinos in thermal leptogenesis~\cite{fukugita1986,sakharov1967,harvey1990,khlebnikov1988,covi1996,buchmuller1998,buchmuller2005,davidson2008,donia2014}. In the type-I seesaw~\cite{minkowski1977,gellmann1979,yanagida1980,mohapatra1980}, the measured light-neutrino masses do not determine the high-energy Yukawa couplings, and the final baryon-to-photon ratio $\eta_B$ depends on the evolution of lepton flavour. Charged-lepton Yukawa interactions gradually resolve the lepton states produced in the decays, spectator reactions redistribute the charges, and washout acts on their flavour composition~\cite{barbieri2000,nardi2006higgs,nardi2006,abada2006,blanchet2007,desimone2007,spectator2001,beneke2011,dev2014,moffat2018,cirigliano2010,frossard2013,devreview2018,garbrecht2014,jukkala2021}. The leptons produced in a decay are in a superposition of charged-lepton flavours. While this superposition persists, the lepton flavour density matrix has off-diagonal entries in the charged-lepton mass basis, which we call flavour coherences. They encode the interference between flavours, which a description in terms of flavour populations alone cannot capture. We refer to the damping of these coherences by charged-lepton Yukawa interactions as decoherence. A long-lived, nearly decoupled heavy neutrino can also delay flavour equilibration by modifying the expansion history~\cite{roshanShow2026}.

Asymmetries produced by the heavier neutrinos can survive the washout of the lightest heavy neutrino, denoted by $N_1$, if part of them lies in flavour directions that $N_1$ does not wash out~\cite{dibari2005,engelhard2007,bertuzzo2011,paul2026memory,refiorentin2015,dibari2015,dibari2016,dibari2017,hambye2012,buchmuller2002}. In this paper, the parent of a lepton asymmetry is the heavy neutrino $N_i$ whose CP-violating decays produce it, and the parent-aligned direction is the lepton flavour direction into which $N_i$ decays. Refs.~\cite{antusch2012,blanchet2013} showed that compensating flavour asymmetries (phantom terms) can contribute to the final baryon asymmetry when subsequent washout acts differently on the flavours. Ref.~\cite{blanchet2013} formulated density-matrix equations for arbitrary heavy-neutrino masses, derived multistage Boltzmann equations for hierarchical masses in well-defined flavour regimes, and showed that interference terms in the parent-aligned basis are washed out at half the rate of the total asymmetry in the coherent single-parent limit. More recently, Ref.~\cite{yu2026} decomposed the dominant nonthermal $N_2$ CP source into a projector-aligned and a trace-free part and showed that the latter governs the sign reversal along a two-parent nonthermal leptogenesis trajectory as the timing of washout changes relative to decoherence. The flavour-covariant description of lepton charges in Ref.~\cite{fong2022} was extended to all five Standard Model fermion charge matrices in Ref.~\cite{fong2026}.

Refs.~\cite{antusch2012,blanchet2013,yu2026} established that phantom terms and trace-free CP sources can contribute to the final asymmetry and even reverse its sign. They did not address, in thermal leptogenesis with three hierarchical heavy neutrinos, whether large cancellations occur predominantly within or between parents, how large the trace-free response is relative to the combined heavier-neutrino response across seesaw parameter points, or how the loss of this source information affects the accuracy of approximate flavour equations. 

\begin{table*}[t]
\caption{Main notation and abbreviations.}
\label{tab:notation}
\begin{ruledtabular}
\setlength{\tabcolsep}{4pt}
\noindent\makebox[\linewidth][c]{%
\resizebox{1.00\linewidth}{!}{%
\begin{tabular}{lp{0.36\textwidth}lp{0.35\textwidth}}
Symbol & Meaning & Symbol or label & Meaning\\
\hline
\multicolumn{2}{l}{\textit{Parameters and transport}} & \multicolumn{2}{l}{\textit{Responses and cancellation diagnostics}}\\
$N_i,M_i$ & Heavy parent neutrino and its mass & $R_{i,a},R_i$ & Signed block response and parent total\\
$\alpha,\beta$ & Charged-lepton flavour indices & $R_i^{\rm d},R_i^{\rm o}$ & Diagonal and coherent response sums\\
$T,z,z_i,r_i$ & Temperature; $M_1/T$, $M_i/T$, $M_i/M_1$ & $R_i^\parallel,R_i^\perp$ & Aligned and trace-free source responses\\
$Y_\nu,y_i,h$ & Neutrino Yukawa matrix, its $i$th column, $Y_\nu^\dagger Y_\nu$ & $\Lambda$ & Sum of absolute block responses, $\sum_{i,a}|R_{i,a}|$\\
$Y_{N_i},Y_{N_i}^{\rm eq}$ & Parent yield and equilibrium yield & $\kappa_{18}$ & Total cancellation, $\Lambda/|\eta_B|$\\
$K_i,D_i$ & Decay parameter and decay coefficient & $\kappa_{\rm within}$ & Within-parent cancellation, $\Lambda/\sum_i|R_i|$\\
$\overline P_i,\Pi_\alpha$ & Parent washout and charged-flavour projectors & $\kappa_{\rm between}$ & Between-parent cancellation, $\sum_i|R_i|/|\eta_B|$\\
$\varepsilon^{(i)},\varepsilon_i$ & CP-asymmetry matrix and its trace & $\kappa_i^{\rm flav}$ & Cancellation among blocks of parent $i$\\
$\varepsilon_{i,a}$ & Diagonal or Hermitian coherent CP block (block index $a$) & $\kappa_{{\rm src},i}$ & Cancellation of parent-projected source blocks\\
$\varepsilon^{(i)}_\parallel,\varepsilon^{(i)}_\perp$ & Projector-aligned and trace-free interference CP terms & $\kappa_{\rm d/o},\kappa_{\rm d},\kappa_{\rm o}$ & Between diagonal/coherent sums; within diagonal and coherent blocks, respectively\\
$Y_{\Delta X}$ & Charge matrix of $X=\ell,E,Q,U,D$ & $\kappa^{\rm RP}_{18}$ & Total cancellation from RP-$T$ responses\\
$q,A,b_i$ & Charge vector, transport operator, parent source & \multicolumn{2}{l}{\textit{Approximate transport (Sec.~\ref{sec:approx})}}\\
$G_i,W_i$ & Washout coefficient and $G_i/(4Y_{\rm nor})$ & RP-$T$, RP-$M_1$ & Five-matrix regime projection, regime at $T$ or fixed at $T=M_1$\\
$\gamma_E$ & Charged-lepton Yukawa transfer rate & SB-$T$, SB-$M_1$ & Scalar Boltzmann equations, regime at $T$ or fixed at $T=M_1$\\
$Y_{\Delta_\alpha},Y_{B-L}$ & Flavour $B/3-L_\alpha$ and total $B-L$ yields & SB-proj & Scalar Boltzmann equations with heavier-parent overlap projection\\
$\eta_B,\eta_B^{\rm obs}$ & Final and observed baryon-to-photon ratios & LDM & Lepton density matrix with full CP sources\\
$\eta_B^{\rm full},\eta_B^{\rm approx}$ & Full and approximate transport results & LDM-pop & Lepton density matrix with reconstructed population sources\\
$\Omega,f$ & Complex orthogonal seesaw matrix and mass rescaling factor & $C^{(n)},C=1$ & Equilibrium spectator matrix; omission of spectators\\
$\Delta_i$ & Orthogonal projection of diagonal CP source in Eq.~(\ref{eq:delta}) & $T_\tau,T_\mu,J_n$ & Flavour thresholds and regime projector\\
$\eta_\parallel,\eta_\perp$ & Toy-model aligned and unwashed efficiencies & $\epsilon_{ia},P^0_{ia},p_i^a$ & Sector CP asymmetry, branching ratio and aligned overlap weight\\
$F_i$ & Linear map from parent CP source to final response & $\rho^{(i)},X_i$ & Parent lepton density matrix and its CP source\\
\end{tabular}%
}
}
\end{ruledtabular}
\end{table*}

In this study, we address these questions by resolving the final asymmetry into 18 signed block responses across sampled seesaw parameter points. We split the CP-asymmetry matrix of each heavy neutrino into three diagonal and three off-diagonal Hermitian blocks and evolve each block separately in the five-matrix Standard Model transport system of Ref.~\cite{fong2026}, using the public code \texttt{BOLEH}~\cite{bolehsoftware} and its Standard Model rate parametrizations. 

Our two main results are as follows. First, in the charged-lepton mass basis, cancellation among the flavour blocks of individual parents dominates over cancellation between parents. Second, the trace-free interference term makes a large contribution independently of the choice of flavour-block partition: the median signed ratio of its response to the combined heavier-neutrino response is 84\% in points where that response exceeds 20\% of the final asymmetry in magnitude. The novelty lies in resolving and quantifying these cancellations and contributions, not in the source decomposition itself.

Section~\ref{sec:framework} gives the transport equations, the source decomposition, and the seesaw parameter points. The numerical results are presented in Sec.~\ref{sec:locate}, and their origin is analysed with a single-parent model in Sec.~\ref{sec:toy}. The partition-independent contribution of the trace-free part is evaluated in Sec.~\ref{sec:perp}. Section~\ref{sec:approx} defines the approximate flavour equations and compares their accuracy with the full transport, and Sec.~\ref{sec:conclusions} gives our conclusions. The dependence on the calibration and sampling is examined in Appendix~\ref{sec:robustness}, and that on the reaction rates and the flavour basis in Appendix~\ref{sec:rates}. Appendix~\ref{sec:numerical} gives the numerical prescription.

For ease of reference, Table~\ref{tab:notation} collects the main symbols and abbreviations used below.

\section{Framework}\label{sec:framework}

We assume radiation domination throughout. We denote the heavy neutrinos by $N_i$ with $i=1,2,3$ and masses $M_1<M_2<M_3$. We work in the basis in which the charged-lepton Yukawa matrix and the heavy Majorana mass matrix $M_R={\rm diag}(M_1,M_2,M_3)$ are diagonal, with flavour indices $\alpha,\beta=e,\mu,\tau$. The neutrino Yukawa matrix is $Y_\nu=M_D/v$, where $M_D$ is the Dirac mass matrix and $v=174\,{\rm GeV}$ is the Higgs vacuum expectation value. Its elements $Y_{\alpha i}\equiv(Y_\nu)_{\alpha i}$ couple lepton flavour $\alpha$ to $N_i$. We write $h\equiv Y_\nu^\dagger Y_\nu$, with $h_{ij}=\sum_\alpha Y^*_{\alpha i}Y_{\alpha j}$, and the light-neutrino mass matrix is $M_\nu=-v^2Y_\nu M_R^{-1}Y_\nu^T$. The mass splittings of the heavy neutrinos are large compared with their widths, so the decays lie outside the resonant regime~\cite{pilaftsis1997,pilaftsis2004,devresonant2018}.

Let $T$ be the plasma temperature, $z=M_1/T$, $r_i=M_i/M_1$, $z_i=r_iz$, and $Y_{N_i}=n_{N_i}/s$ the number density $n_{N_i}$ normalized to the entropy density $s$, with $g_*=106.75$ relativistic degrees of freedom. With two spin degrees of freedom, the Maxwell--Boltzmann equilibrium yield is $Y^{\rm eq}_{N_i}=(45/2\pi^4g_*)z_i^2{\rm K}_2(z_i)$, where ${\rm K}_n$ is the modified Bessel function of the second kind. We use the effective neutrino mass $\tilde m_i=v^2h_{ii}/M_i$, the equilibrium neutrino mass $m_*=1.08\times10^{-3}\,{\rm eV}$, the decay parameter $K_i=\tilde m_i/m_*$, and $D_i=K_ir_i^2z\,{\rm K}_1(z_i)/{\rm K}_2(z_i)$ for the decay coefficient. We use the Hubble rate $H_{\rm rad}=1.66\sqrt{g_*}T^2/M_{\rm Pl}$ and entropy density $s=(2\pi^2/45)g_*T^3$, with Planck mass $M_{\rm Pl}=1.2209\times10^{19}\,{\rm GeV}$. The top-Yukawa ($H$) and gauge-mediated ($A$) $\Delta L=1$ reaction densities (where $L$ in $\Delta L$ is lepton number), normalized by $sH_{\rm rad}z$, are denoted by $\hat\gamma_{H_{s,t},i}$ and $\hat\gamma_{A_{s,t},i}$ for the $s$ and $t$ channels, respectively. The abundances obey~\cite{buchmuller2005,davidson2008,giudice2004,strumia2006,fongnardi2012}
\begin{equation}
\begin{aligned}
\frac{dY_{N_i}}{dz}={}&-\Bigl[D_i+\frac{2(\hat\gamma_{H_s,i}+\hat\gamma_{A_s,i})}{Y^{\rm eq}_{N_i}}\\
&+\frac{4(\hat\gamma_{H_t,i}+\hat\gamma_{A_t,i})}{Y^{\rm eq}_{N_i}}\Bigr]\bigl(Y_{N_i}-Y^{\rm eq}_{N_i}\bigr).
\end{aligned}
\label{eq:heavy}
\end{equation}
The heavy neutrinos start from zero abundance at $z_{\rm in}=10^{-3}$.

The flavour direction of the leptons produced by $N_i$ is described by a rank-one projector~\cite{beneke2011,moffat2018}. With $y_i$ the $i$th column of $Y_\nu$, in the index convention of the charge variables used below, the projector reads
\begin{equation}
\overline P_i=\frac{y_i^*y_i^T}{y_i^\dagger y_i},\qquad y_i=(Y_{ei},Y_{\mu i},Y_{\tau i})^T,
\label{eq:projector}
\end{equation}
the complex conjugate of the usual $y_iy_i^\dagger/(y_i^\dagger y_i)$. We denote by $\varepsilon^{(i)}_{\rm M}$ the Hermitian CP-asymmetry matrix of $N_i$ in the conventional lepton-flavour convention, and by $\varepsilon^{(i)}=(\varepsilon^{(i)}_{\rm M})^*$ its complex conjugate in the charge-matrix convention of Eq.~(\ref{eq:projector}). With $j=1,2,3$, $x_{ji}=M_j^2/M_i^2$, and the loop functions of a dimensionless argument $x$,
\begin{equation}
\begin{aligned}
f_1(x)&=\frac{2\sqrt{x}}{3}\left[(1+x)\ln\frac{1+x}{x}-\frac{2-x}{1-x}\right],\\
f_2(x)&=\frac{2}{3(x-1)},
\end{aligned}
\label{eq:loop-functions}
\end{equation}
the non-resonant CP-asymmetry matrix in manifestly Hermitian form has entries~\cite{covi1996,buchmuller1998,moffat2018}
\begin{equation}
\begin{aligned}
\varepsilon^{(i)}_{{\rm M},\alpha\beta}=\frac{3}{32\pi h_{ii}}\sum_{j\neq i}\Bigl\{&i\bigl[Y_{\alpha i}Y^*_{\beta j}h_{ji}-Y_{\alpha j}Y^*_{\beta i}h_{ij}\bigr]f_1(x_{ji})\\
+\,&i\bigl[Y_{\alpha i}Y^*_{\beta j}h_{ij}-Y_{\alpha j}Y^*_{\beta i}h_{ji}\bigr]f_2(x_{ji})\Bigr\}.
\end{aligned}
\label{eq:epsilon}
\end{equation}
The second Yukawa product in each bracket is the Hermitian conjugate of the first (with the same loop function).

The charges of the left-handed leptons $\ell$, right-handed charged leptons $E$, quark doublets $Q$, and right-handed up- and down-type quarks $U$, $D$ are Hermitian $3\times3$ matrices $Y_{\Delta X}$ ($X=\ell,E,Q,U,D$), expressing particle--antiparticle charge asymmetries normalized to $s$~\cite{fong2026,bolehsoftware}. For the quark matrices the three generation indices are ordered analogously to $e,\mu,\tau$. For any Hermitian $3\times3$ matrix $X$ (here a matrix rather than a particle label), we use nine real coordinates,
\begin{equation}
\begin{aligned}
{\rm vec}(X)=\bigl(&X_{ee},X_{\mu\mu},X_{\tau\tau},{\rm Re}\,X_{e\mu},{\rm Im}\,X_{e\mu},\\
&{\rm Re}\,X_{e\tau},{\rm Im}\,X_{e\tau},{\rm Re}\,X_{\mu\tau},{\rm Im}\,X_{\mu\tau}\bigr)^T,
\end{aligned}
\label{eq:hermitian-vector}
\end{equation}
and collect the 45 coordinates in the column vector
\begin{equation}
q=\begin{pmatrix}{\rm vec}(Y_{\Delta\ell})\\{\rm vec}(Y_{\Delta E})\\{\rm vec}(Y_{\Delta Q})\\{\rm vec}(Y_{\Delta U})\\{\rm vec}(Y_{\Delta D})\end{pmatrix},
\label{eq:q-vector}
\end{equation}
so that, for example, $q_1=(Y_{\Delta\ell})_{ee}$, $q_5={\rm Im}\,(Y_{\Delta\ell})_{e\mu}$, and $q_{19}$ is the first-generation diagonal entry of $Y_{\Delta Q}$. With $0_{36}$ denoting the 36 vanishing source coordinates of the other four charge matrices, the source of parent $i$ is~\cite{giudice2004,moffat2018}
\begin{equation}
b_i(z)=\Bigl({\rm vec}\bigl[\varepsilon^{(i)}\bigr]D_i\bigl(Y_{N_i}-Y^{\rm eq}_{N_i}\bigr),\,0_{36}\Bigr)^T.
\label{eq:source}
\end{equation}
We retain the vacuum decay CP asymmetry and include $\Delta L=1$ scatterings in the abundance and washout equations, but not their CP-violating source terms. This is a decay-source approximation, not a complete treatment of the scattering contributions. Ref.~\cite{nardiracker2007} showed that these sources can compensate early and late asymmetry production for a vanishing initial heavy-neutrino abundance, particularly in weak washout.
For fixed abundances from Eq.~(\ref{eq:heavy}), the charges obey the linear system~\cite{fong2026,bolehsoftware}
\begin{equation}
\frac{dq}{dz}=A(z)\,q+\sum_{i=1}^3b_i(z),
\label{eq:linear}
\end{equation}
where $A$ is the $45\times45$ matrix of the linear map $q\mapsto dq/dz$ given by the right-hand side without the sources, with components
\begin{equation}
A_{mn}(z)=\frac{\partial}{\partial q_n}\left(\frac{dq_m}{dz}\right).
\label{eq:A-components}
\end{equation}
The operator $A$ contains inverse-decay and $\Delta L=1$ washout, charged-lepton and quark Yukawa transfer, QCD and electroweak sphalerons, and a low-temperature $\Delta L=2$ term; for the Standard Model charge transfer and sphalerons we use the equations and rate parametrizations of Refs.~\cite{fong2026,bolehsoftware}. With the Higgs charge $Y_{\Delta H}$ fixed by hypercharge neutrality and the multiplicities and statistical factors $(g_\ell,\zeta_\ell,g_H,\zeta_H)=(2,1,2,2)$, where $g_\ell$ and $g_H$ are the $SU(2)_L$ multiplicities of the lepton and Higgs doublets and $\zeta_\ell$ and $\zeta_H$ are the statistical factors relating a charge asymmetry to its chemical potential for a fermion and a boson, respectively, we use the rescaled charges $\tilde Y_\ell=Y_{\Delta\ell}/(g_\ell\zeta_\ell)$ and $\tilde Y_H=Y_{\Delta H}/(g_H\zeta_H)$. With these charges and the normalization $Y_{\rm nor}=15/(8\pi^2g_*)$, the washout coefficient of parent $i$ is
\begin{equation}
G_i=D_iY^{\rm eq}_{N_i}+4(\hat\gamma_{H_t,i}+\hat\gamma_{A_t,i})+2(\hat\gamma_{H_s,i}+\hat\gamma_{A_s,i})\frac{Y_{N_i}}{Y^{\rm eq}_{N_i}},
\label{eq:washout-coefficient}
\end{equation}
With $\{X_1,X_2\}=X_1X_2+X_2X_1$ denoting the anticommutator of two matrices, the washout enters the lepton equation as
\begin{equation}
\left.\frac{dY_{\Delta\ell}}{dz}\right|_{\rm wash}=-\sum_i\frac{G_i}{2Y_{\rm nor}}\left[\frac12\bigl\{\overline P_i,\tilde Y_\ell\bigr\}+\overline P_i\tilde Y_H\right].
\label{eq:washout}
\end{equation}
For $z\geq5$ the off-shell $\Delta L=2$ washout of the Weinberg operator is added~\cite{beneke2011,frossard2013,blanchet2012}. We use $c_\nu=Y_\nu^*M_R^{-1}Y_\nu^\dagger$, the complex conjugate of the Weinberg-operator coefficient as required by the convention of Eq.~(\ref{eq:projector}), and $F_{\rm QP}=1.0347$, the ratio of the thermal quasiparticle phase-space integral to its massless Maxwell--Boltzmann value. The washout term reads
\begin{align}
\left.\frac{dY_{\Delta\ell}}{dz}\right|_{\Delta L=2}={}&-\frac{F_{\rm QP}T^3}{\pi^3H_{\rm rad}z}\Bigl[\frac14\bigl(2c_\nu\tilde Y_\ell^Tc_\nu^\dagger\\
&\quad+c_\nu c_\nu^\dagger\tilde Y_\ell+\tilde Y_\ell c_\nu c_\nu^\dagger\bigr)+\frac12\tilde Y_Hc_\nu c_\nu^\dagger\Bigr].\nonumber
\label{eq:weinberg}
\end{align}
With the baryon yield $Y_B=\frac13{\rm Tr}(Y_{\Delta Q}+Y_{\Delta U}+Y_{\Delta D})$, the flavour-lepton yield $Y_{L_\alpha}=(Y_{\Delta\ell})_{\alpha\alpha}+(Y_{\Delta E})_{\alpha\alpha}$, $Y_{\Delta_\alpha}=Y_B/3-Y_{L_\alpha}$, and $Y_{B-L}=\sum_\alpha Y_{\Delta_\alpha}$, we evaluate the baryon-to-photon ratio at $z_f=40$. We use the charged-lepton Yukawa eigenvalues $(y_{e_e},y_{e_\mu},y_{e_\tau})=(2.94\times10^{-6},6.07\times10^{-4},1.02\times10^{-2})$ (also held fixed in the high-temperature charge transfer) and conversion coefficients $C_{\rm sph}=0.3328$ and $F_{\rm sph}=0.0279$ in the flavour-dependent sphaleron freeze-out relation of Ref.~\cite{benso2026},
\begin{equation}
\eta_B=7.039\left[C_{\rm sph}Y_{B-L}+F_{\rm sph}\sum_\alpha y_{e_\alpha}^2Y_{\Delta_\alpha}\right]_{z_f},
\label{eq:terminal}
\end{equation}
where the factor 7.039 is the present entropy-to-photon ratio~\cite{harvey1990,khlebnikov1988,davidson2008,fong2026,donia2014,babudibari2024,benso2026}. The charges vanish at $z_{\rm in}$. We assume the individual $Y_{\Delta_\alpha}$ are frozen after $z_f$, so Eq.~(\ref{eq:terminal}) describes their eventual conversion at the electroweak crossover, not the baryon abundance at $T=M_1/z_f$.

For the lepton-number-violating rates we use fixed representative values of the $SU(2)_L$, $U(1)_Y$, and $SU(3)_c$ gauge couplings $g_2,g_Y,g_3$, the top Yukawa coupling $y_t$, and the Higgs quartic coupling $\lambda$: $(g_2,g_Y,g_3,y_t,\lambda)=(0.55,0.42,0.80,0.50,0.10)$. The $\Delta L=1$ scattering rates use the thermal-mass-regulated cross sections of Ref.~\cite{giudice2004}. For the top-Yukawa channels, let $I_{s,t}(z_i)$ be the dimensionless thermal integrals defined in Eq.~(\ref{eq:thermal-integrals}). The rates are
\begin{equation}
\hat\gamma_{H_{s,t},i}=\frac{45M_{\rm Pl}h_{ii}y_t^2I_{s,t}(z_i)}{128\pi^6\times1.66\,g_*^{3/2}M_1}.
\label{eq:thermal-rates}
\end{equation}
The gauge-mediated rates have the same normalization without $y_t^2$. The thermal mass ratios are fixed at $T=M_i$ for the top-Yukawa channels, whereas their temperature dependence is retained for the gauge-mediated channels. Appendix~\ref{sec:numerical} specifies the thermal kernels and numerical prescription.

With $|\alpha\rangle$ the unit vector for charged-lepton flavour $\alpha$ and $\Pi_\alpha=|\alpha\rangle\langle\alpha|$ its projector, each CP matrix is split into three diagonal and three Hermitian coherent blocks~\cite{beneke2011,moffat2018},
\begin{equation}
\begin{aligned}
\varepsilon_{i,\alpha\alpha}&=\Pi_\alpha\varepsilon^{(i)}\Pi_\alpha,\\
\varepsilon_{i,\alpha\beta}&=\Pi_\alpha\varepsilon^{(i)}\Pi_\beta+\Pi_\beta\varepsilon^{(i)}\Pi_\alpha\quad(\alpha<\beta),
\end{aligned}
\label{eq:blocks}
\end{equation}
labelled $a=ee,\mu\mu,\tau\tau,e\mu,e\tau,\mu\tau$, where $\alpha<\beta$ refers to the ordering $e,\mu,\tau$. The off-diagonal blocks are called coherent blocks because they source the flavour coherences of $Y_{\Delta\ell}$. Let $q_{i,a}$ solve Eq.~(\ref{eq:linear}) with only block $a$ of parent $i$ as source. By linearity~\cite{beneke2011,dev2014,fong2026} $q=\sum_{i,a}q_{i,a}$, and the signed response of block $(i,a)$ is the value $R_{i,a}$ of Eq.~(\ref{eq:terminal}) evaluated with $q_{i,a}$,
\begin{equation}
R_i=\sum_aR_{i,a},\qquad \eta_B=\sum_{i,a}R_{i,a}=\sum_iR_i.
\label{eq:responses}
\end{equation}
The coherent blocks are traceless and contribute to $\eta_B$ only through flavour-dependent transport. Changing the source partition changes its block responses but not $R_i$ or $\eta_B$.

We quantify the cancellation using the sum of the absolute block responses $\Lambda=\sum_{i,a}|R_{i,a}|$:
\begin{equation}
\begin{aligned}
\kappa_{18}&=\frac{\Lambda}{|\eta_B|},\qquad
\kappa_{\rm within}=\frac{\Lambda}{\sum_i|R_i|},\\
\kappa_{\rm between}&=\frac{\sum_i|R_i|}{|\eta_B|}.
\end{aligned}
\label{eq:kappa}
\end{equation}
Thus, $\kappa_{18}=\kappa_{\rm within}\kappa_{\rm between}$. For nonzero denominators, each factor equals one without cancellation at its level and grows with the cancellation: $\kappa_{\rm within}$ measures the cancellation among the blocks of the same parent and $\kappa_{\rm between}$ that among the parent totals. For each parent we also use
\begin{equation}
\kappa_i^{\rm flav}=\frac{\sum_a|R_{i,a}|}{|R_i|},
\label{eq:kappa-parent}
\end{equation}
which measures the cancellation among the six block responses of $N_i$. The factor $\kappa_{\rm between}$ depends only on the parent totals and is partition invariant; $\kappa_{\rm within}$ and $\kappa^{\rm flav}_i$ are not. 

In the seesaw parametrization of Ref.~\cite{casas2001}, let $D_m={\rm diag}(m_1,m_2,m_3)$, where $m_i$ are the light-neutrino masses. We use $\Omega=O_{23}(\omega_3)O_{13}(\omega_2)O_{12}(\omega_1)$, $\Omega\Omega^T=1$, where $O_{jk}$ is a rotation in the $jk$ plane and $\omega_n$ are complex angles ($n=1,2,3$). The PMNS matrix $U$ is in the standard convention, with mixing angles $\theta_{ij}$, Dirac phase $\delta_{\rm CP}$, and Majorana phases $\alpha_{21},\alpha_{31}$~\cite{pontecorvo1958,maki1962,davidson2008,nufit60journal,nufit61release}. The Dirac mass matrix is then $M_D=iU^*D_m^{1/2}\Omega M_R^{1/2}$. We adopt normal ordering, favoured but not established by recent oscillation and cosmological analyses~\cite{nufit60journal,nufit61release,t2k2025,desi2025,ivanov2026}, with the lightest mass $m_1=0$. With $\Delta m^2_{ij}=m_i^2-m_j^2$, the oscillation parameters are fixed to the NuFIT 6.1 best fit with IceCube IC24 and Super-Kamiokande atmospheric data~\cite{nufit60journal,nufit61release}: $\theta_{12}=33.76^\circ$, $\theta_{13}=8.62^\circ$, $\theta_{23}=43.29^\circ$, $\Delta m^2_{21}=7.537\times10^{-5}\,{\rm eV}^2$, $\Delta m^2_{31}=2.511\times10^{-3}\,{\rm eV}^2$, and $\delta_{\rm CP}=212^\circ$. The light masses are then $(m_1,m_2,m_3)=(0,0.0087,0.0501)\,{\rm eV}$ with sum $0.0588\,{\rm eV}$. Cosmological mass bounds are model dependent~\cite{desi2025,ivanov2026}.

The two possible parameter choices $\Omega$ and $\Omega^*$ are called CP branches here: they have opposite signs for the imaginary parts of the complex angles, with the low-energy mixing parameters kept fixed. The Majorana phases $\alpha_{21}$ and $\alpha_{31}$ are uniform in $[0,2\pi)$, the real and imaginary parts of the three complex angles $\omega_n$ are uniform in $[-\pi,\pi]$ and $[-0.45,0.45]$, and the proposal value $M_1^{\rm prop}$ of the lightest heavy-neutrino mass is log-uniform in $[10^9,10^{13}]\,{\rm GeV}$. We use the mass ratios $M_1:M_2:M_3=1:4:20$, $1:3:10$, and $1:10:50$ with 67, 67, and 66 points, labelled $k=0,\dots,199$ in this order, and require the largest singular value $\sigma_{\max}(Y_\nu)<1$. Each point is calibrated to $\eta_B^{\rm obs}=(6.120\pm0.038)\times10^{-10}$~\cite{yeh2026,planck2018} by a common positive rescaling factor $f$, with $M_i\to fM_i$, $Y_\nu\to\sqrt fY_\nu$, which preserves $M_\nu$ (Appendix~\ref{sec:robustness}). The scale factors range from $1.843\times10^{-3}$ to $1.823\times10^{2}$ (median 0.7142), and the $\Omega^*$ branch is used for 98 points. All 200 points lie within 0.320 standard deviations of the observed value, with $\sigma_{\max}(Y_\nu)<0.2456$ and $M_1$ between $4.46\times10^{9}\,{\rm GeV}$ and $2.61\times10^{12}\,{\rm GeV}$. All three heavy neutrinos have nonzero Yukawa couplings in every point.

\section{Flavour cancellation}\label{sec:locate}

\begin{figure*}[t]
\includegraphics[width=\textwidth]{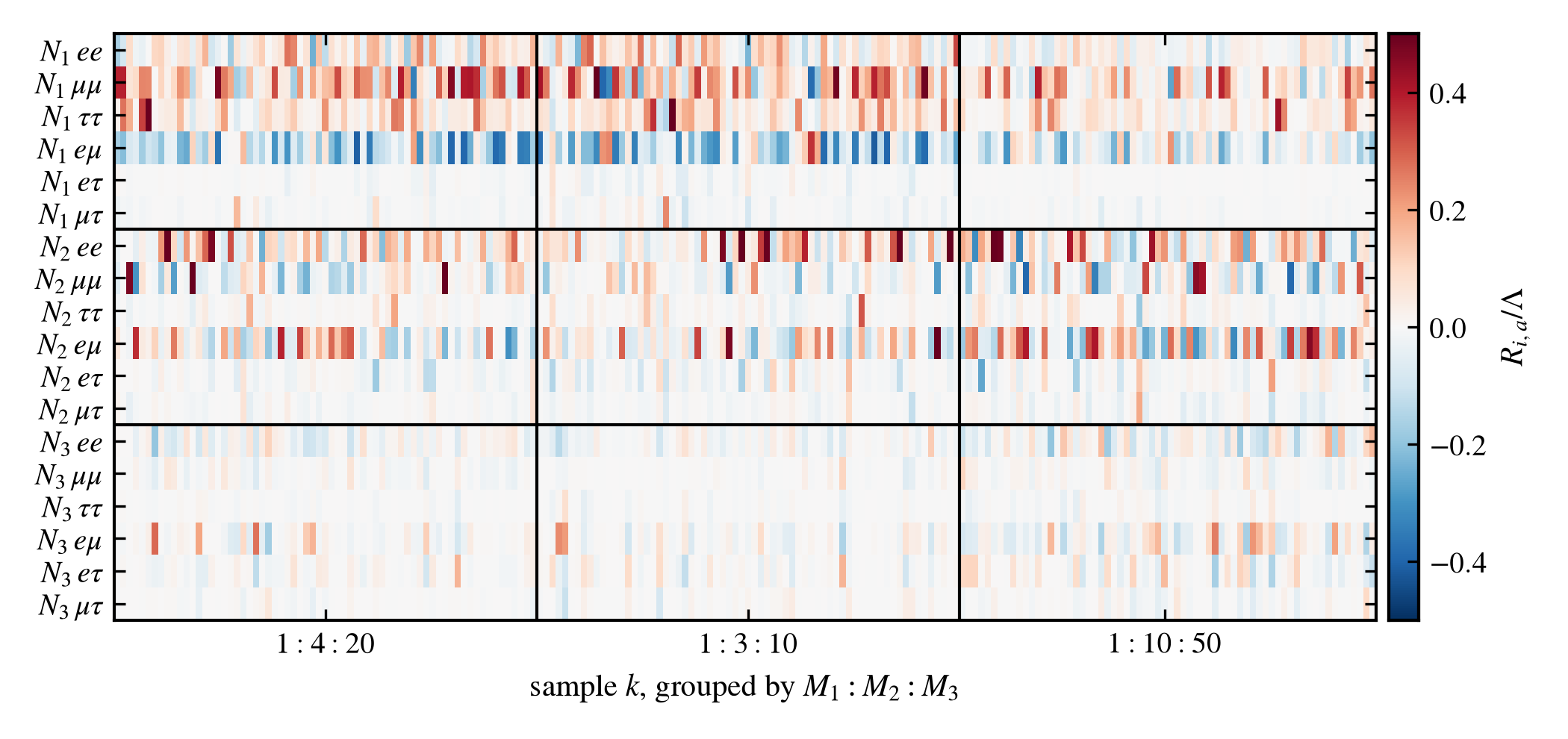}
\caption{Signed responses $R_{i,a}/\Lambda$ for the 200 points, grouped by mass ratio and parent. Opposite signs within each parent reveal the flavour cancellation.}
\label{fig:channelmap}
\end{figure*}

We first examine the cancellation among the 18 signed responses shown in Fig.~\ref{fig:channelmap}. Positive and negative responses occur within each parent. The total cancellation factor $\kappa_{18}$ is of order ten, with medians 6.98, 7.78, and 10.09 for the three mass ratios.

\begin{figure}[t]
\includegraphics[width=\columnwidth]{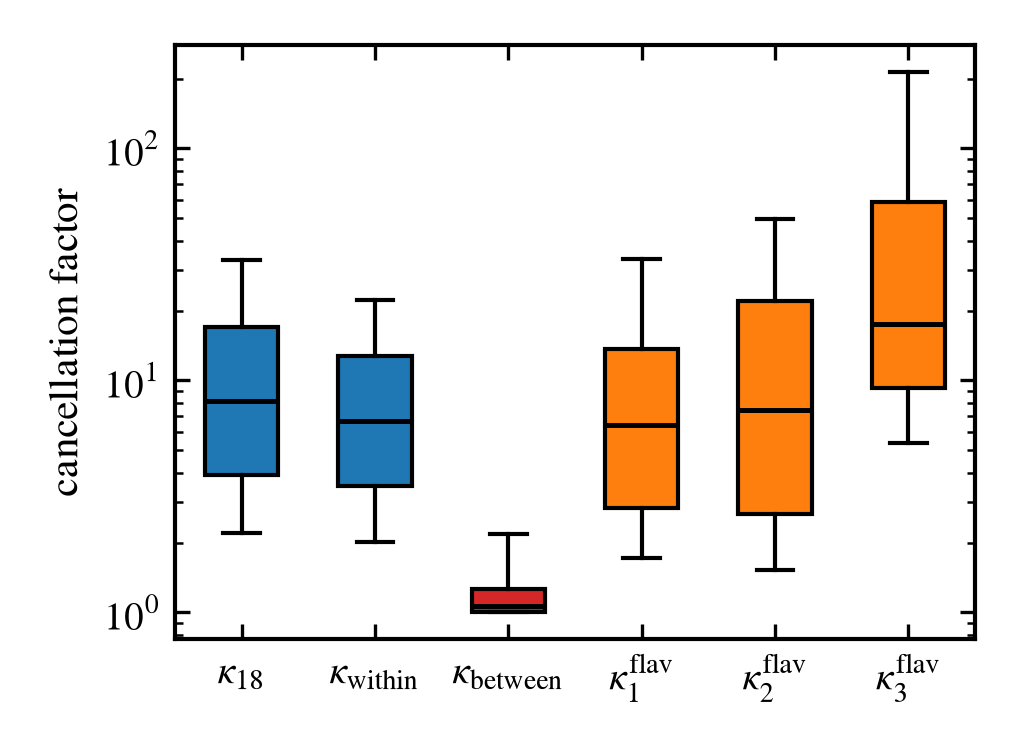}
\caption{Cancellation factors for the 200 points; boxes and whiskers denote the interquartile and 10th--90th percentile ranges. Within-parent cancellation dominates.}
\label{fig:kappa}
\end{figure}

Table~\ref{tab:kappa} and Fig.~\ref{fig:kappa} distinguish the two levels of cancellation. The factor $\kappa_{\rm within}$ follows $\kappa_{18}$, while $\kappa_{\rm between}$ stays close to one, and $\kappa_{\rm within}>\kappa_{\rm between}$ in 196 of 200 points. The small $\kappa_{\rm between}$ is independent of source partition. Appendix~\ref{sec:robustness} examines calibration and random-sign controls.

\begin{table}[t]
\caption{Cancellation factors of Eqs.~(\ref{eq:kappa}) and~(\ref{eq:kappa-parent}) for the 200 points (medians and 10th--90th percentiles). The parent totals add almost without mutual cancellation, but the flavour blocks of each parent cancel strongly.}
\label{tab:kappa}
\begin{ruledtabular}
\begin{tabular}{lcc}
Factor & Median & 10th--90th percentile\\
\hline
$\kappa_{18}$ & 8.17 & 2.19--33.68\\
$\kappa_{\rm within}$ & 6.66 & 2.01--22.63\\
$\kappa_{\rm between}$ & 1.06 & 1.00--2.18\\
$\kappa_1^{\rm flav}$ & 6.42 & 1.71--33.50\\
$\kappa_2^{\rm flav}$ & 7.44 & 1.52--49.68\\
$\kappa_3^{\rm flav}$ & 17.56 & 5.36--222.84
\end{tabular}
\end{ruledtabular}
\end{table}

\begin{figure}[t]
\includegraphics[width=\columnwidth]{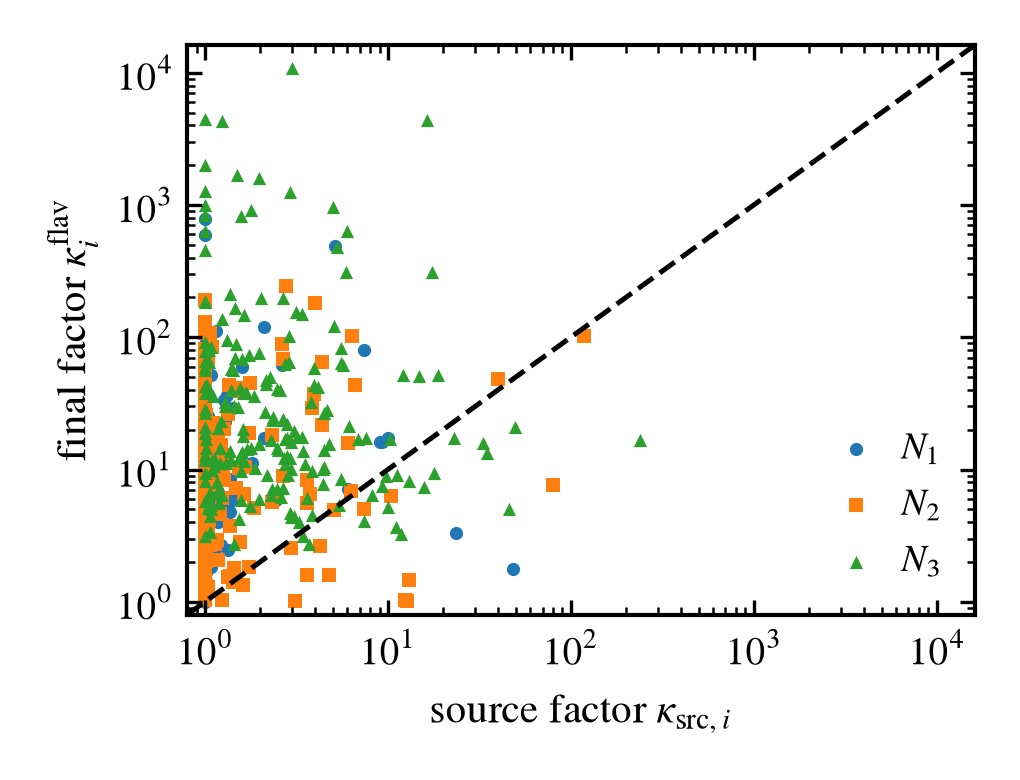}
\caption{Source and final cancellation factors for each parent; the dashed line denotes equality. The larger final factors indicate cancellation generated during transport.}
\label{fig:source}
\end{figure}

To determine whether the cancellation is already present in the source or develops during transport, we project each source block on the washout direction of its parent and measure the cancellation among the projected blocks,
\begin{equation}
\kappa_{{\rm src},i}=\frac{\sum_a|{\rm Tr}(\overline P_i\varepsilon_{i,a})|}{\bigl|\sum_a{\rm Tr}(\overline P_i\varepsilon_{i,a})\bigr|}.
\label{eq:source-factor}
\end{equation}
The decay profile cancels in this ratio. The projected sources of $N_1$ and $N_2$ mostly have a common sign, and even for $N_3$ the source factor is far smaller than $\kappa_3^{\rm flav}$ (Tables~\ref{tab:kappa} and~\ref{tab:diagnostics}). As shown in Fig.~\ref{fig:source}, the final factor exceeds the source factor in 198, 184, and 180 out of 200 points for $N_1$, $N_2$, and $N_3$, respectively.

To identify the transport process responsible, we compare two controls with the full source matrices kept fixed. Without transport ($A=0$) the coherent blocks have no response, and the diagonal blocks give $R_{i,\alpha\alpha}\propto(\varepsilon^{(i)})_{\alpha\alpha}$. Removing only the inverse-decay, $\Delta L=1$, and $\Delta L=2$ washout from $A$, while keeping Yukawa transfer and sphalerons, gives the same within-parent factors to $3.3\times10^{-6}$ point by point. In both controls $\kappa_{\rm within}$ stays close to one (Table~\ref{tab:diagnostics}) and exceeds $\kappa_{\rm between}$ in only 58 of 200 points. In the absence of lepton-number-violating washout, Standard Model charge transfer and sphalerons conserve each $Y_{\Delta_\alpha}$ apart from the specified CP injection. The comparison therefore identifies flavour-dependent washout as the origin of the large cancellation within this fixed-source model. The washout-off calculation is used only as a diagnostic within the fixed-source approximation.

We next distinguish the types of flavour contributions that cancel. Grouping the responses of each parent into diagonal and coherent sums, $R_i^{\rm d}=R_{i,ee}+R_{i,\mu\mu}+R_{i,\tau\tau}$ and $R_i^{\rm o}=R_{i,e\mu}+R_{i,e\tau}+R_{i,\mu\tau}$, we define
\begin{equation}
\begin{aligned}
\kappa_{\rm d/o}&=\frac{\sum_i\bigl(|R_i^{\rm d}|+|R_i^{\rm o}|\bigr)}{\sum_i|R_i|},\qquad
\kappa_{\rm d}=\frac{\sum_i\sum_{\alpha}|R_{i,\alpha\alpha}|}{\sum_i|R_i^{\rm d}|},\\
\kappa_{\rm o}&=\frac{\sum_i\sum_{\alpha<\beta}|R_{i,\alpha\beta}|}{\sum_i|R_i^{\rm o}|}.
\end{aligned}
\label{eq:kappa-do}
\end{equation}
Table~\ref{tab:diagnostics} shows that $\kappa_{\rm d/o}$ is large, while $\kappa_{\rm d}$ and $\kappa_{\rm o}$ stay close to one, and $\kappa_{\rm d/o}>\kappa_{\rm between}$ in 193 of 200 points. In the charged-lepton basis the cancellation is between the diagonal and the coherent responses of the same parent. Keeping only the 15 diagonal charge coordinates and the nine diagonal source blocks, i.e., applying three-flavour equations at all temperatures, removes this cancellation (median within-parent factor 1.09) and reverses the sign of $\eta_B$ in 13.5\% of the points.

\begin{figure}[t]
\includegraphics[width=\columnwidth]{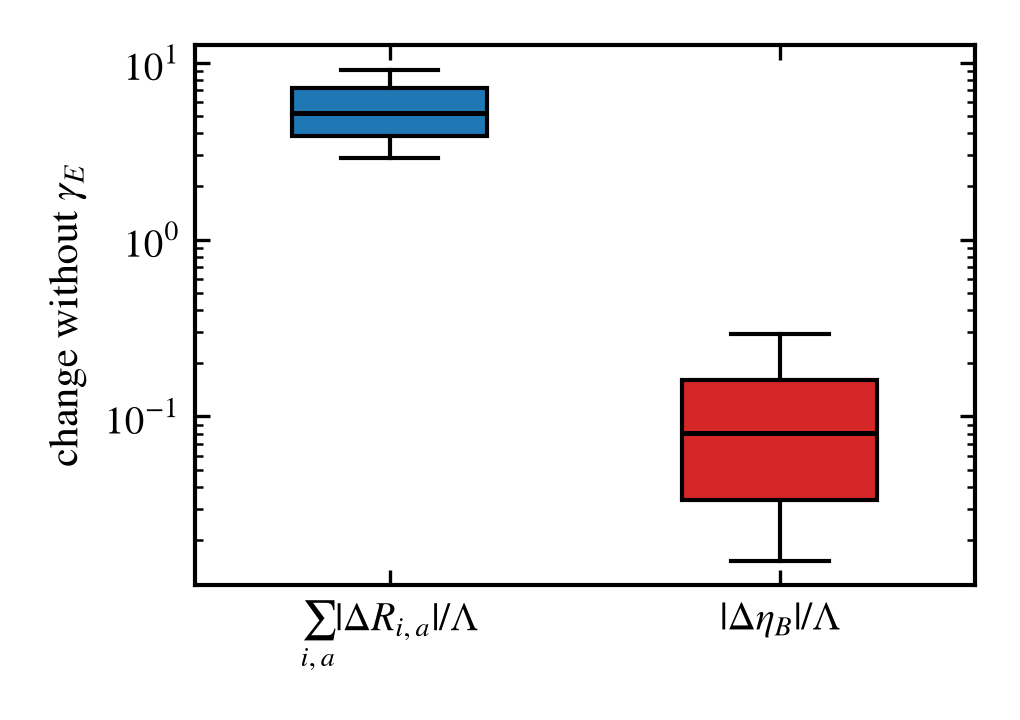}
\caption{Changes in the block responses and $\eta_B$ on removing charged-lepton Yukawa transfer, in units of $\Lambda$ (boxes and whiskers as in Fig.~\ref{fig:kappa}). The individual responses change far more than their sum.}
\label{fig:charged-lepton}
\end{figure}

To examine the role of charged-lepton decoherence, we now vary the charged-lepton Yukawa transfer. In comparisons below, $\Delta R_{i,a}$ and $\Delta\eta_B$ denote the changes from the full transport result at the same seesaw point. Removing only the charged-lepton Yukawa transfer term from $A$ (with rate $\gamma_E(z)$) changes the 18 responses by a median $\sum_{i,a}|\Delta R_{i,a}|=5.17\,\Lambda$ and reverses the sign of at least one response with $|R_{i,a}|\geq0.01\Lambda$ in 64.5\% of the points, while $\eta_B$ changes by a median of only 8.1\% of $\Lambda$ (Fig.~\ref{fig:charged-lepton}). At the same time $\kappa_{18}$ grows by about an order of magnitude (Table~\ref{tab:diagnostics}). 

\begin{table}[t]
\caption{Source and response cancellation factors (medians and 10th--90th percentiles). The cancellation is weak without washout and grows without charged-lepton transfer.}
\label{tab:diagnostics}
\begin{ruledtabular}
\begin{tabular}{lcc}
Factor & Median & 10th--90th percentile\\
\hline
$\kappa_{{\rm src},1}$ & 1.00 & 1.00--1.33\\
$\kappa_{{\rm src},2}$ & 1.00 & 1.00--3.77\\
$\kappa_{{\rm src},3}$ & 1.98 & 1.00--9.32\\
$\kappa_{\rm d/o}$ & 4.89 & 1.59--16.34\\
$\kappa_{\rm d}$ & 1.22 & 1.02--2.20\\
$\kappa_{\rm o}$ & 1.14 & 1.01--2.00\\
\hline
$\kappa_{\rm within}$, no washout & 1.24 & 1.02--2.40\\
$\kappa_{18}$, no $\gamma_E$ & 73.42 & 15.71--290.55
\end{tabular}
\end{ruledtabular}
\end{table}

\begin{figure}[t]
\includegraphics[width=\columnwidth]{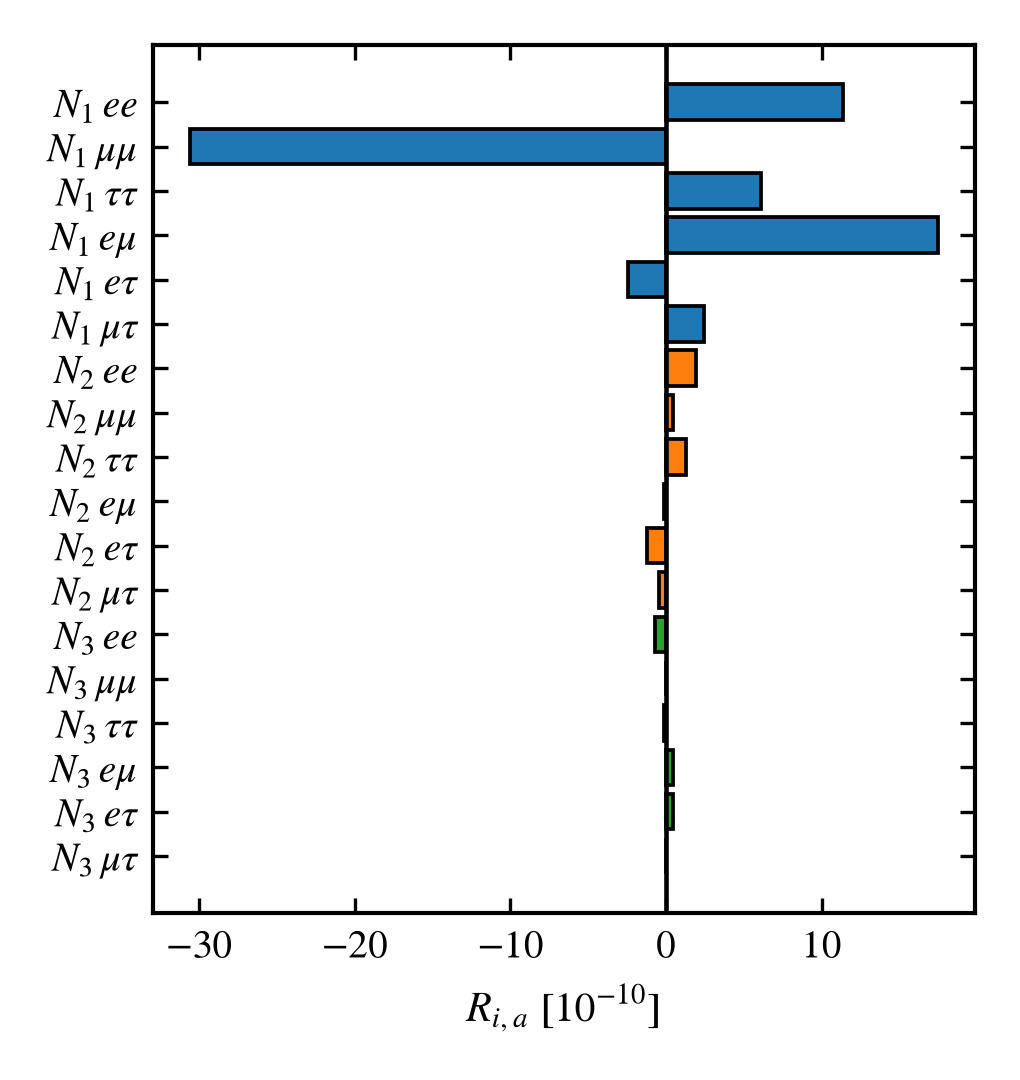}
\caption{Block responses at point $k=77$ in units of $10^{-10}$. The three parent totals are positive despite large opposing contributions within $N_1$.}
\label{fig:benchmark}
\end{figure}

As an illustration of the cancellation in the full calculation, point $k=77$ (mass ratio $1:3:10$, scale factor $f=0.610498$) has $M_1=1.0278\times10^{11}\,{\rm GeV}$, $\sigma_{\max}(Y_\nu)=0.0273$, and $\eta_B=6.1240\times10^{-10}$. The parent totals are $+4.23$, $+1.80$, and $+0.09$ (units of $10^{-10}$), and the sums of the magnitudes of their block responses are 70.34, 5.46, and 1.74. In Fig.~\ref{fig:benchmark} the negative $N_1\,\mu\mu$ response ($-30.58\times10^{-10}$) is almost cancelled by the positive $N_1\,e\mu$, $N_1\,ee$, and $N_1\,\tau\tau$ responses, while the three parent totals are positive, so that $\kappa_{\rm within}=12.66$ and $\kappa_{\rm between}=1.00$. The net asymmetry is thus much smaller than the individual flavour responses of $N_1$.

\section{Origin of the cancellation}\label{sec:toy}

The opposing diagonal and coherent responses can be understood from the structure of the CP matrix. Every term of Eq.~(\ref{eq:epsilon}) carries the Yukawa vector of the decaying parent on the left or on the right. In the convention of Eq.~(\ref{eq:projector}) this implies the exact identity
\begin{equation}
\bigl(1-\overline P_i\bigr)\varepsilon^{(i)}\bigl(1-\overline P_i\bigr)=0.
\label{eq:identity}
\end{equation}
This is the projector form of the CP-matrix structure in Eq.~(66) of Ref.~\cite{blanchet2013}; the corresponding projector identities for $N_2$ were also given in Appendix B of Ref.~\cite{yu2026}. It follows that ${\rm Tr}\,\varepsilon^{(i)}={\rm Tr}(\overline P_i\varepsilon^{(i)})\equiv\varepsilon_i$, so the trace resides in the washout-aligned component. For the split into the diagonal part $\varepsilon^{(i)}_{\rm d}$ and the coherent part $\varepsilon^{(i)}_{\rm o}$ (the sums of the diagonal and coherent blocks of Eq.~(\ref{eq:blocks}), respectively),
\begin{equation}
\Delta_i\equiv{\rm Tr}\bigl[(1-\overline P_i)\varepsilon^{(i)}_{\rm d}\bigr]=-{\rm Tr}\bigl[(1-\overline P_i)\varepsilon^{(i)}_{\rm o}\bigr].
\label{eq:delta}
\end{equation}
The diagonal and coherent sums have opposite projections onto the flavour directions orthogonal to their parent. These projections cancel in the unsplit source.

To see how these source projections give rise to cancellation during transport, consider one parent of mass $M$ with $z=M/T$, decay parameter $K$, abundance $Y_N$, lepton charge matrix $Y$, normalized flavour state $|y\rangle$ with $\langle y|y\rangle=1$, and projector $\overline P=|y\rangle\langle y|$. In this model $z_{\rm in}$ and $z_f$ denote the initial and final values of $z$, and $Y(z_{\rm in})=0$. We normalize $Y_N$ to the relativistic equilibrium abundance, so that $Y_N^{\rm eq}(z)=z^2{\rm K}_2(z)/2$ and $Y_N^{\rm eq}(0)=1$. We use the decay rate $D(z)=Kz\,{\rm K}_1(z)/{\rm K}_2(z)$ and source profile $S(z)=D(z)(Y_N-Y_N^{\rm eq})$. The inverse-decay washout rate is $W(z)=\frac14Kz^3{\rm K}_1(z)$, the injected CP matrix (or one of its blocks) is $X$, and $\Gamma(z)$, which describes the charged-lepton Yukawa interactions, damps the off-diagonal part $Y^{\rm off}$ of $Y$ in the flavour basis. The evolution is
\begin{equation}
\frac{dY}{dz}=S(z)X-\frac{W(z)}2\bigl\{\overline P,Y\bigr\}-\Gamma(z)Y^{\rm off}.
\label{eq:toy}
\end{equation}
The observable is $-c_B\,{\rm Tr}Y(z_f)$ with $c_B>0$ collecting the flavour-independent conversion factors of Eq.~(\ref{eq:terminal}); the small flavour-dependent correction is omitted in this toy model.

For $\Gamma=0$ the components of $Y$ along $|y\rangle\langle y|$, $|y\rangle\langle y_\perp|$, and $|y_\perp\rangle\langle y_\perp|$, with a normalized orthogonal state $|y_\perp\rangle$ satisfying $\langle y|y_\perp\rangle=0$, decay at the rates $W$, $W/2$, and 0. With the usual efficiency $\eta_\parallel=\int dz\,S(z)\exp[-\int_z^{z_f}W(z')\,dz']$ and the efficiency $\eta_\perp=\int dz\,S(z)$ of a component that is not washed out, where the outer integrals run from $z_{\rm in}$ to $z_f$, the final trace $F[X]$ for an injected $X$ is
\begin{equation}
F[X]=\eta_\perp{\rm Tr}X-(\eta_\perp-\eta_\parallel){\rm Tr}\bigl(\overline PX\bigr).
\label{eq:toy-response}
\end{equation}
In the coherent single-parent limit, the washout operator in Eq.~(\ref{eq:toy}) agrees with Eq.~(65) of Ref.~\cite{blanchet2013}, up to conventions, and reproduces the half-rate damping of the parent-aligned interference terms in Eqs.~(66) and (67) of that reference. These terms have zero trace, so their half-rate efficiency does not appear in Eq.~(\ref{eq:toy-response}).

For decays and inverse decays only, $S=-dY_N/dz$ and hence $\eta_\perp=Y_N(z_{\rm in})-Y_N(z_f)$. It approaches one for a thermal initial abundance at sufficiently small $z_{\rm in}$ when the parent has completely decayed, and vanishes for a zero initial abundance when none remain. For a thermal initial abundance in strong washout, $|\eta_\parallel|\ll|\eta_\perp|$. For zero initial abundance, the decay-only single-parent model instead has $\eta_\perp\simeq0$; neither the thermal hierarchy nor its sign prediction follows in that limit. In the full system, scattering also changes $Y_{N_i}$, and washout by the other parents is not aligned with $\overline P_i$. Applying Eq.~(\ref{eq:toy-response}) to the diagonal and coherent parts $X^{\rm d}$ and $X^{\rm o}$ of the CP matrix, with the total CP asymmetry $\varepsilon={\rm Tr}X={\rm Tr}X^{\rm d}$, ${\rm Tr}X^{\rm o}=0$, and the single-parent projection $\Delta={\rm Tr}[(1-\overline P)X^{\rm d}]={\rm Tr}(\overline PX^{\rm o})$ from Eq.~(\ref{eq:delta}), gives the two responses $R^{\rm d}$ and $R^{\rm o}$, defined by multiplying the corresponding final traces by $-c_B$:
\begin{equation}
\begin{aligned}
R^{\rm d}&=-c_B\bigl[\eta_\parallel\varepsilon+(\eta_\perp-\eta_\parallel)\Delta\bigr],\\
R^{\rm o}&=c_B(\eta_\perp-\eta_\parallel)\Delta,\qquad
R^{\rm d}+R^{\rm o}=-c_B\eta_\parallel\varepsilon.
\end{aligned}
\label{eq:toy-blocks}
\end{equation}
The diagonal and coherent sums each contain a projected charge in the directions not washed out by their parent, with efficiency $\eta_\perp$. This is a property of the separate block responses, not an additional protected component of the physical unsplit source. These charges are equal and opposite by Eq.~(\ref{eq:identity}), leaving the washed-out parallel contribution to the trace. This resembles the protected components of Refs.~\cite{barbieri2000,dibari2005,engelhard2007}, but here the two separately propagated blocks have opposite protected parts and their sum has none.

For a single parent, the factor of Eq.~(\ref{eq:kappa-do}) is $\kappa_{\rm d/o}=(|R^{\rm d}|+|R^{\rm o}|)/|R^{\rm d}+R^{\rm o}|$. With $\phi=(\eta_\perp/\eta_\parallel-1)\Delta/\varepsilon$, Eq.~(\ref{eq:toy-blocks}) gives
\begin{equation}
\kappa_{\rm d/o}=|1+\phi|+|\phi|.
\label{eq:toy-kappa}
\end{equation}
The blocks have opposite signs whenever $\phi>0$ or $\phi<-1$, in particular whenever $|\phi|\gg1$. For $|\eta_\perp/\eta_\parallel|\gg1$ and $|\phi|\gg1$, $\kappa_{\rm d/o}\simeq2|\eta_\perp/\eta_\parallel||\Delta/\varepsilon|$. For thermal initial abundance the inverse parallel efficiency grows with $K$~\cite{buchmuller2005}. Without washout $\eta_\parallel=\eta_\perp$, so $R^{\rm o}=0$ and $\kappa_{\rm d/o}=1$ if the net response is nonzero, as in the washout-off control of Sec.~\ref{sec:locate}. For the washout-aligned split into $\varepsilon\overline P$ and $X^\perp=X-\varepsilon\overline P$ both ${\rm Tr}X^\perp$ and ${\rm Tr}(\overline PX^\perp)$ vanish, so $F[X^\perp]=0$ and this split shows no cancellation. For a diagonal/coherent partition in another flavour basis, the same formulas hold with $\Delta$ evaluated in that basis: the dynamical factor $\eta_\perp/\eta_\parallel$ is common to all partitions, the geometric factor is not.

We now restore decoherence. Let $p_\alpha=|\langle\alpha|y\rangle|^2$ be the branching fraction into flavour $\alpha$. For $\Gamma\to\infty$ the coherent injection is removed immediately, so $R^{\rm o}\to0$, $R^{\rm d}$ becomes the flavoured result with washout rates $p_\alpha W$, and $\kappa_{\rm d/o}=1$. This is analogous to phantom flavour terms~\cite{antusch2012,blanchet2013}, although the split-source diagnostic is not identical to those terms; in particular, the latter can undergo washout at production. For fast but finite decoherence in the two-flavour model, when $\Gamma$ is large compared with $W$ and the inverse variation scales of the source and damping profile, the coherence injected by the coherent block follows the source, $Y^{\rm off}\simeq SX^{\rm o}/\Gamma$, and feeds the diagonal charges through the anticommutator. To first order in $W/\Gamma$, for two flavours labelled $a,b$ with branching fractions $p_c=|\langle c|y\rangle|^2$ ($c=a,b$) and survival factors $E_c(z)=\exp[-p_c\int_z^{z_f}W(z')\,dz']$,
\begin{equation}
R^{\rm o}\simeq c_B\frac{\Delta}{2}\int dz\,S(z)\frac{W(z)}{\Gamma(z)}\bigl[E_a(z)+E_b(z)\bigr].
\label{eq:toy-fast}
\end{equation}
The integral runs from $z_{\rm in}$ to $z_f$. The expansion assumes the fast-damping condition where the source contributes appreciably. For the thermal initial abundance used below the integrand has a fixed sign apart from $\Delta$, and the response is suppressed by $W/\Gamma$ with the same sign as the coherent limit. Ref.~\cite{yu2026} showed, for nonthermal $N_2$ production, that the timing of $N_1$ washout relative to $e$--$\mu$ decoherence controls the propagated contribution of the trace-free part of the $N_2$ source, in line with this picture.

\begin{figure}[t]
\includegraphics[width=\columnwidth]{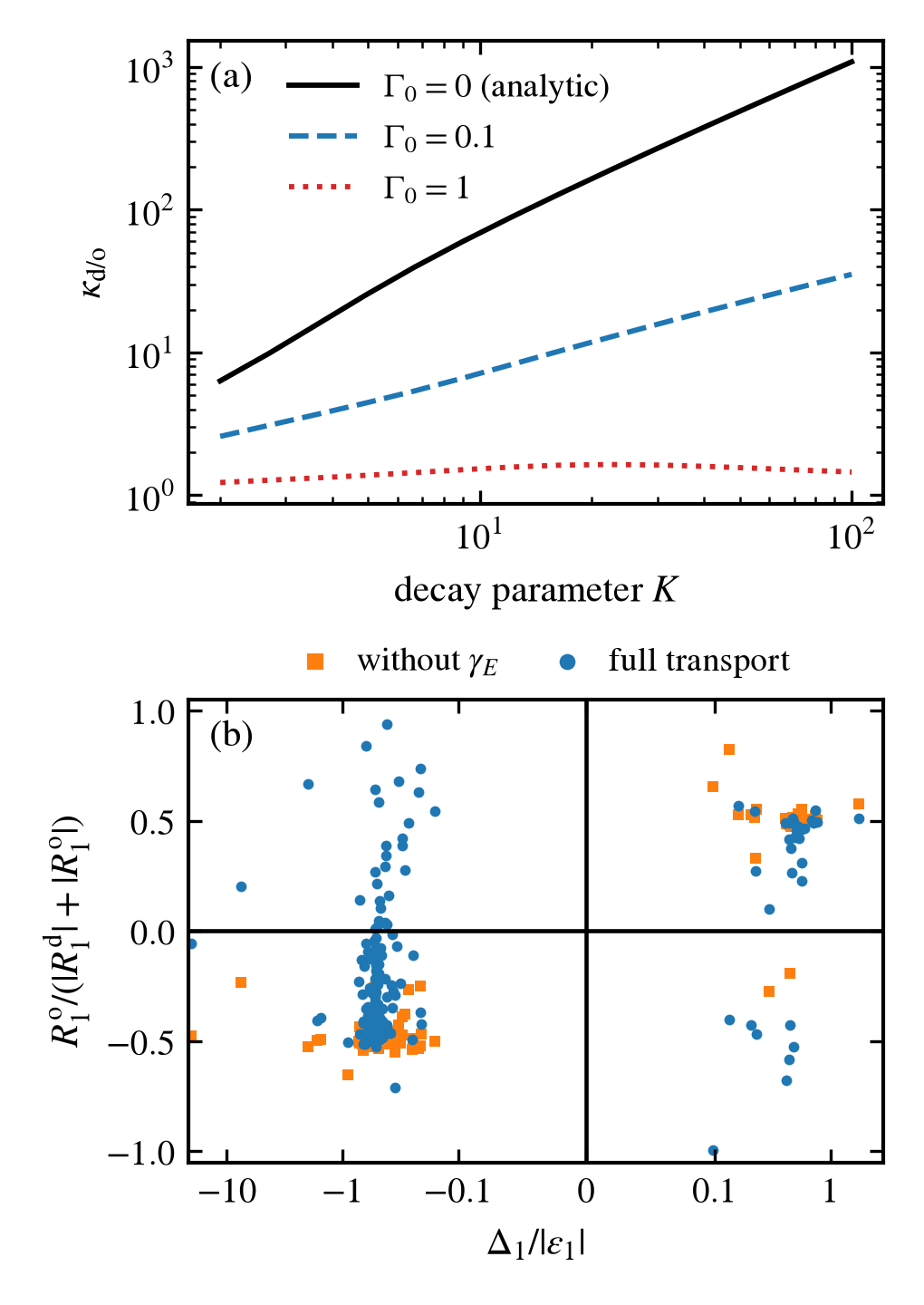}
\caption{(a) Two-flavour cancellation factor versus $K$ for several damping strengths $\Gamma_0$. (b) Coherent response $R_1^{\rm o}/(|R_1^{\rm d}|+|R_1^{\rm o}|)$ versus $\Delta_1/|\varepsilon_1|$ in the full calculation and without $\gamma_E$ (linear scale within $\pm0.1$). Decoherence suppresses the opposing response.}
\label{fig:toy}
\end{figure}

Table~\ref{tab:toy} and Fig.~\ref{fig:toy}a show the two-flavour solutions from $z=10^{-3}$ to $z_f=60$ with a thermal initial abundance and damping rate $\Gamma(z)=\Gamma_0z$, where $\Gamma_0$ is a constant. The coherent and fast-decoherence limits agree with Eqs.~(\ref{eq:toy-kappa}) and~(\ref{eq:toy-fast}) (Appendix~\ref{sec:rates}). As $K$ increases, the coherent cancellation grows while decoherence suppresses it. Turning to the full transport, the coherent responses in the first and third quadrants of Fig.~\ref{fig:toy}b have the sign of $\Delta_1$, as predicted by Eq.~(\ref{eq:toy-blocks}) when $\eta_\perp-\eta_\parallel>0$. Values near $\pm\frac12$ indicate nearly equal and opposite diagonal and coherent responses; removing charged-lepton transfer brings the points closer to this pattern.

\begin{table}[t]
\caption{Two-flavour model of Eq.~(\ref{eq:toy}) with $|y\rangle=(\cos0.6,\sin0.6)^T$, $|y_\perp\rangle=(-\sin0.6,\cos0.6)^T$, and $X=\overline P+2(|y\rangle\langle y_\perp|+|y_\perp\rangle\langle y|)$, so that $\varepsilon=1$ and $\Delta/\varepsilon=1.11$: parallel efficiency and $\kappa_{\rm d/o}$ for several decoherence strengths $\Gamma_0$. Columns labelled $\Gamma_0$ give the cancellation factors, while $\eta_\parallel$ measures survival along the washout direction. Without decoherence the cancellation grows as the inverse washout efficiency; moderate decoherence removes most of it.}
\label{tab:toy}
\begin{ruledtabular}
\begin{tabular}{lccccc}
$K$ & $\eta_\parallel$ & $\Gamma_0=0$ & $\Gamma_0=0.1$ & $\Gamma_0=1$ & $\Gamma_0=10$\\
\hline
5 & 0.082 & 25.8 & 4.47 & 1.38 & 1.04\\
20 & 0.013 & 165 & 11.8 & 1.64 & 1.04\\
50 & 0.0045 & 490 & 22.3 & 1.56 & 1.03\\
\end{tabular}
\end{ruledtabular}
\end{table}

For all three parents, agreement with this conditional sign prediction improves without charged-lepton transfer (Table~\ref{tab:toy-tests}). For $M_1$ expressed in GeV, in the intervals $M_1<10^{10}$, $10^{10}\leq M_1<10^{11}$, $10^{11}\leq M_1<10^{12}$, and $M_1\geq10^{12}$, the median values of $\kappa_{\rm within}$ are 2.58, 4.95, 12.88, and 64.79, respectively (11, 128, 54, and 7 points). This mass dependence in the sampled ensemble is consistent with weaker decoherence at higher temperature, although the mass bins do not hold the washout parameters or the source geometry fixed. The estimate $-(R_1^{\rm o}/R_1)/(\Delta_1/\varepsilon_1)$ of $\eta_\perp/\eta_\parallel-1$ is positive in 78\% of the points. A thermal initial abundance raises the median $\kappa_{\rm within}$ from 6.66 to 7.93. Splitting $\varepsilon^{(i)}$ into the terms proportional to the loop functions $f_1$ and $f_2$, we find that the traceless $f_2$ piece has a larger cancellation factor than the $f_1$ piece in 87\% of the points (medians 13.83 and 6.46, respectively).

\begin{table}[t]
\caption{Fractions of the 200 points displaying the sign pattern of Eq.~(\ref{eq:toy-blocks}) for $\eta_\perp-\eta_\parallel>0$ and the opposite-sign response pattern. Agreement improves without charged-lepton transfer.}
\label{tab:toy-tests}
\begin{ruledtabular}
\begin{tabular}{lcc}
Prediction & Full transport & Without $\gamma_E$\\
\hline
${\rm sign}\,R^{\rm o}_1={\rm sign}\,\Delta_1$ & 82\% & 99\%\\
${\rm sign}\,R^{\rm o}_2={\rm sign}\,\Delta_2$ & 70\% & 92\%\\
${\rm sign}\,R^{\rm o}_3={\rm sign}\,\Delta_3$ & 59\% & 78\%\\
$R^{\rm d}_iR^{\rm o}_i<0$, all parents & 89.0\% & 98.5\%
\end{tabular}
\end{ruledtabular}
\end{table}

\section{Partition-independent orthogonal response}\label{sec:perp}

The cancellation discussed in Secs.~\ref{sec:locate} and~\ref{sec:toy} is between responses that depend on the partition of the CP source. The parent totals $R_i$, however, are independent of this partition. The identity~(\ref{eq:identity}) gives a partition that is fixed by the CP matrix itself, $\varepsilon^{(i)}=\varepsilon^{(i)}_\parallel+\varepsilon^{(i)}_\perp$ with
\begin{equation}
\varepsilon^{(i)}_\parallel={\rm Tr}\bigl(\overline P_i\varepsilon^{(i)}\bigr)\overline P_i,\qquad
\varepsilon^{(i)}_\perp=\varepsilon^{(i)}-\varepsilon^{(i)}_\parallel.
\label{eq:aligned-split}
\end{equation}
By Eq.~(\ref{eq:identity}), $\varepsilon^{(i)}_\parallel$ contains the trace $\varepsilon_i$, and $\varepsilon^{(i)}_\perp=\overline P_i\varepsilon^{(i)}(1-\overline P_i)+(1-\overline P_i)\varepsilon^{(i)}\overline P_i$ is trace free and satisfies ${\rm Tr}(\overline P_i\varepsilon^{(i)}_\perp)=0$. It is an interference term, not a source supported entirely in the protected orthogonal flavour directions. The split is independent of the choice of flavour basis.

At fixed heavy-neutrino histories, let $F_i$ be the linear map from a parent-$i$ source matrix to its contribution to $\eta_B$. For every decomposition $\varepsilon^{(i)}=\sum_a\varepsilon_{i,a}$,
\begin{equation}
\begin{aligned}
R_i&=F_i\bigl[\varepsilon^{(i)}\bigr]=\sum_aF_i[\varepsilon_{i,a}],\\
R_i^\parallel&=F_i\bigl[\varepsilon^{(i)}_\parallel\bigr],\qquad
R_i^\perp=F_i\bigl[\varepsilon^{(i)}_\perp\bigr].
\end{aligned}
\label{eq:invariant}
\end{equation}
Here $R_i=R_i^\parallel+R_i^\perp$, with $R_i^\parallel$ the response to the CP asymmetry along the washout direction. Unlike the cancellation factors of the individual flavour blocks, $R_i^\perp$ is independent of the source partition. It vanishes for coherent single-parent washout, since both ${\rm Tr}\,\varepsilon^{(i)}_\perp$ and ${\rm Tr}(\overline P_i\varepsilon^{(i)}_\perp)$ vanish in Eq.~(\ref{eq:toy-response}). In the full calculation this interference can feed the trace through charged-lepton decoherence followed by washout, through washout by other parents, or through the off-shell $\Delta L=2$ operator. The flavour-dependent term of Eq.~(\ref{eq:terminal}) also permits a small response even when the final $B-L$ trace vanishes. The washout-aligned split shows little cancellation (median factor 1.29; Table~\ref{tab:partitions}).

\begin{table}[t]
\caption{Partition-independent orthogonal responses of Eq.~(\ref{eq:invariant}) in the full transport (medians and 10th--90th percentiles of signed ratios). The fourth row uses the 145 points in which $|R_2+R_3|>0.2|\eta_B|$. The trace-free term is important for the heavier-neutrino contributions in the median, with large point-to-point variations.}
\label{tab:perp}
\begin{ruledtabular}
\begin{tabular}{lcc}
Ratio & Median & 10th--90th percentile\\
\hline
$R_1^\perp/R_1$ & 10\% & $-92$--$98$\%\\
$R_2^\perp/R_2$ & 83\% & $-25$--$173$\%\\
$R_3^\perp/R_3$ & 113\% & $-74$--$341$\%\\
$(R_2^\perp+R_3^\perp)/(R_2+R_3)$ & 84\% & $-9$--$169$\%\\
$\sum_iR_i^\perp/\eta_B$ & 38\% & $-62$--$145$\%
\end{tabular}
\end{ruledtabular}
\end{table}

Table~\ref{tab:perp} quantifies the contribution of the trace-free part to the final asymmetry. The median of $R_2^\perp/R_2$ is 83\%, against 10\% for $N_1$, so the $N_2$ contribution depends mainly on the flavour structure of its CP matrix and not only on its trace. Summed over the three parents, $\sum_iR_i^\perp/\eta_B$ has a median of 38\% and a 10th--90th percentile range from $-62$\% to $145$\%; its magnitude exceeds 20\% in 79\% of the points.
In the points with a substantial heavier-parent contribution, the orthogonal term gives a median 84\% of $R_2+R_3$, and the parallel responses $R_2^\parallel+R_3^\parallel$ alone reproduce $R_2+R_3$ within a factor of two in 28\% of these points and give the correct sign in 59\%. In some points $R_i^\parallel$ and $R_i^\perp$ have opposite signs, which explains the wide percentile ranges in Table~\ref{tab:perp}. In Sec.~\ref{sec:approx} we use $R_i^\perp$ to test the approximate equations.

\section{Comparison with approximate equations}\label{sec:approx}

We now test whether approximate flavour equations reproduce both the cancellations and the trace-free responses discussed above. Commonly used flavour-regime approximations to Boltzmann transport~\cite{nardi2006,abada2006,blanchet2007,antusch2012,blanchet2013} distinguish three flavour regimes with the approximate boundaries $T_\tau=10^{12}\,{\rm GeV}$ and $T_\mu=10^9\,{\rm GeV}$: above $T_\tau$ the lepton flavours are coherent, between the two temperatures the $\tau$ is resolved while the $e$--$\mu$ sector stays coherent, and below $T_\mu$ all flavours are resolved. The boundaries are approximate; Ref.~\cite{blanchet2007} showed that flavour resolution also depends on the competition with inverse decays.

To compare the flavour approximations on equal footing, we use the same heavy-neutrino histories, decay and $\Delta L=1$ rates, and final baryon conversion. We test regime projection (RP) of the five-matrix system, scalar Boltzmann equations (SB) with equilibrium spectators, and lepton density matrices (LDM) without spectators. The suffix $T$ denotes a temperature-dependent flavour regime, and $M_1$ a regime fixed at $T=M_1$. The equilibrium spectator matrix is denoted by $C$, with $C^{(n)}$ its value for the one-, two-, and three-sector regimes ($n=1,2,3$) and $C=1$ denoting the omission of spectator effects. SB-proj splits each heavier-parent population along and orthogonal to the $N_1$ washout direction in coherent sectors. LDM and LDM-pop evolve lepton density matrices without $\Delta L=2$ washout, using respectively the full CP matrix or a population source reconstructed from flavour-sector traces and branching projectors. We first specify the equations and matching prescriptions and then compare their results.

We denote the baryon-to-photon ratios in an approximate and the full evolution by $\eta_B^{\rm approx}$ and $\eta_B^{\rm full}$, respectively; unqualified $\eta_B$ in the comparisons denotes the full result. We denote the cancellation factor computed from the RP-$T$ responses using Eq.~(\ref{eq:kappa}) by $\kappa^{\rm RP}_{18}$.

For RP, the $45\times45$ projector $J_n$ acts only on the $\ell$ and $E$ charge matrices: $n=1$ retains all entries, $n=2$ removes the $e\tau$ and $\mu\tau$ coherences, and $n=3$ removes all off-diagonal entries. The projected transport is
\begin{equation}
\frac{dq}{dz}=J_nA(z)J_nq+J_n\sum_ib_i(z),
\label{eq:regime}
\end{equation}
with $q\to J_nq$ on entering a new regime. RP-$T$ chooses $n$ at the current temperature, RP-$M_1$ at $T=M_1$, using the thresholds defined above. RP-$M_1$ omits charged-lepton transfer for the 7 points with $M_1>T_\tau$; no calibrated point has $M_1<T_\mu$.

For the SB and LDM equations, the sector index $a$ (distinct from the six-block index of Sec.~\ref{sec:framework}) runs over $e{+}\mu{+}\tau$, $(\gamma,\tau)$ with $\gamma=e{+}\mu$, or $(e,\mu,\tau)$ for $n=1,2,3$. We use the scalar sector CP asymmetries $\epsilon_{ia}=\sum_{\alpha\,\text{in }a}(\varepsilon^{(i)})_{\alpha\alpha}$ and branching ratios $P^0_{ia}=\sum_{\alpha\,\text{in }a}(\overline P_i)_{\alpha\alpha}$, and the washout coefficient $W_j=G_j/(4Y_{\rm nor})$. With the spectator redistribution of Refs.~\cite{nardi2006higgs,nardi2006,antusch2012,devreview2018}, in our convention the lepton and Higgs matrices enter with relative weight $1/2$, giving
\begin{equation}
\begin{aligned}
C^{(2)}&=\frac1{589}\begin{pmatrix}499&-8\\52&502\end{pmatrix},\\
C^{(3)}&=\begin{pmatrix}339/358&6/179&6/179\\6/179&422/537&64/537\\6/179&64/537&422/537\end{pmatrix},
\end{aligned}
\label{eq:coupling}
\end{equation}
and $C^{(1)}=1$. Let $Y_{\Delta_a}^{(i)}$ be the parent-$i$ contribution to the sum of $Y_{\Delta_\alpha}$ in sector $a$, with $b$ running over the same sectors as $a$ and $C_{ab}$ the elements of $C^{(n)}$ in the current regime. The scalar equations are
\begin{equation}
\frac{dY^{(i)}_{\Delta_a}}{dz}=-\epsilon_{ia}D_i\bigl(Y_{N_i}-Y^{\rm eq}_{N_i}\bigr)-\sum_{j=1}^3P^0_{ja}W_j\sum_bC_{ab}Y^{(i)}_{\Delta_b}.
\label{eq:scalar}
\end{equation}
Equation~(\ref{eq:scalar}) follows the scalar transport of Refs.~\cite{nardi2006,abada2006,blanchet2013}. SB-$T$ splits the charges at the thresholds using the parent branching ratios; SB-$M_1$ fixes the regime at $T=M_1$. Unresolved sectors at $z_f$ are split likewise for Eq.~(\ref{eq:terminal}).

The scalar equations neglect coherences and $\Delta L=2$ washout. We compare them with an overlap prescription and with lepton density-matrix evolution. Motivated by Refs.~\cite{dibari2005,engelhard2007,bertuzzo2011,antusch2012}, SB-proj splits the heavier-parent source in each coherent sector into components along and orthogonal to $N_1$. With $y_i^a$ the restriction of $y_i$ to sector $a$, the aligned weight is
\begin{equation}
p^a_i=\frac{\bigl|\langle y_1^a|y_i^a\rangle\bigr|^2}{\langle y_1^a|y_1^a\rangle\langle y_i^a|y_i^a\rangle}.
\label{eq:projection-weight}
\end{equation}
The two source weights are $p_i^a$ and $1-p_i^a$; only the aligned part undergoes $N_1$ washout, while both undergo $N_2$ and $N_3$ washout. Their coherence is neglected. At $T_\tau$ each part is split by its flavour direction and reprojected onto $N_1$ in the remaining coherent sector; at $T_\mu$ it is split into $e$ and $\mu$. For $N_1$ we use SB-$T$.

To isolate the effect of the source approximation within density-matrix evolution, we evolve the $3\times3$ lepton asymmetry density matrix $\rho^{(i)}$, initially zero, without spectators ($C=1$) or $\Delta L=2$ washout. Let $\overline P_i^a$ be the restriction of $\overline P_i$ to sector $a$, normalized to unit trace and embedded in the three-flavour matrix (a sector with zero weight contributes nothing). The source $X_i$ is either $\varepsilon^{(i)}$ restricted to coherent components (LDM) or $\sum_a\epsilon_{ia}\overline P_i^a$ (LDM-pop). The evolution is
\begin{equation}
\frac{d\rho^{(i)}}{dz}=D_i\bigl(Y_{N_i}-Y^{\rm eq}_{N_i}\bigr)X_i-\sum_jW_j\frac12\bigl\{\overline P_j,\rho^{(i)}\bigr\}.
\label{eq:ldm}
\end{equation}
Coherences between resolved flavours are projected out of both the state and its evolution equation, as in Eq.~(\ref{eq:regime}). For the final conversion we identify the parent contribution to $Y_{\Delta_\alpha}$ with $-\rho^{(i)}_{\alpha\alpha}$. In the fully resolved regime this evolution reduces to Eq.~(\ref{eq:scalar}) with $C=1$.

\begin{table*}[!t]
\caption{Comparison with the full calculation: median relative error in $\eta_B$, fractions within 20\%, off by more than a factor of two (Factor 2), and of wrong sign (Sign), and median absolute errors of $R_1$ and $R_2+R_3$ in units of $|\eta_B|$. RP-$T$ is more accurate than the scalar closures; full CP sources improve LDM.}
\label{tab:approx}
\begin{ruledtabular}
\begin{tabular}{lcccccc}
Equations & Median error & Within 20\% & Factor 2 & Sign & $N_1$ & $N_2+N_3$\\
\hline
RP-$T$ & 5.2\% & 79.0\% & 7.0\% & 1.5\% & 0.01 & 0.04\\
RP-$T$, thresholds halved & 4.2\% & 84.5\% & 3.5\% & 1.0\% & -- & --\\
RP-$M_1$ & 5.7\% & 75.0\% & 11.0\% & 3.0\% & -- & --\\
Three flavours at all $T$ & 56.2\% & 21.5\% & 43.5\% & 13.5\% & -- & --\\
Coherent, no $\gamma_E$ & 72.8\% & 15.5\% & 58.0\% & 24.5\% & -- & --\\
\hline
SB-$M_1$ & 55.4\% & 21.5\% & 46.5\% & 14.5\% & 0.16 & 0.50\\
SB-$T$ & 55.4\% & 21.5\% & 46.5\% & 14.5\% & 0.16 & 0.49\\
SB-proj & 58.1\% & 23.5\% & 46.5\% & 16.0\% & 0.16 & 0.48\\
SB-$T$, $C=1$ & 57.9\% & 22.5\% & 49.0\% & 15.0\% & 0.13 & 0.50\\
SB-proj, $C=1$ & 56.8\% & 24.0\% & 46.0\% & 16.0\% & 0.13 & 0.48\\
LDM-pop, $C=1$ & 73.7\% & 15.0\% & 49.5\% & 23.0\% & 0.16 & 0.58\\
LDM, $C=1$ & 16.7\% & 52.5\% & 13.5\% & 4.0\% & 0.07 & 0.18\\
\end{tabular}
\end{ruledtabular}
\end{table*}

\begin{figure}[t]
\includegraphics[width=\columnwidth]{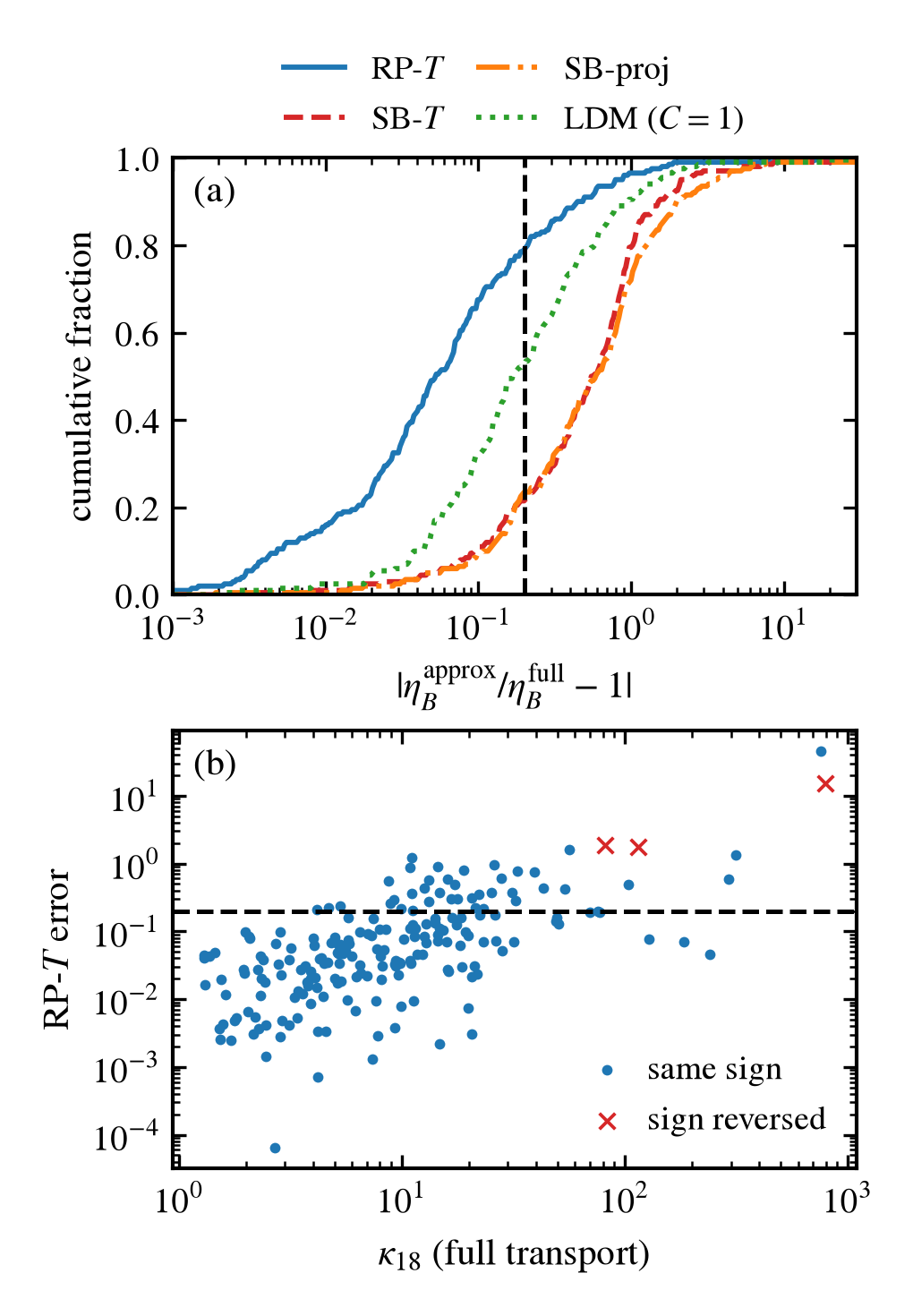}
\caption{(a) Cumulative relative errors $|\eta_B^{\rm approx}/\eta_B^{\rm full}-1|$; the dashed line marks 20\%. The superscripts denote approximate and full evolution, and $C=1$ indicates no spectator redistribution. (b) RP-$T$ error versus the full-transport $\kappa_{18}$; crosses indicate sign reversals. RP-$T$ is more accurate than the scalar prescriptions and degrades at large cancellation.}
\label{fig:approx}
\end{figure}

We first assess RP-$T$, which imposes instantaneous flavour decoherence at the temperature boundaries while retaining the other terms of the five-matrix transport. It has a median relative error $|\eta_B^{\rm approx}/\eta_B^{\rm full}-1|$ of only 5.2\%, with 79.0\% of the points within 20\% (Fig.~\ref{fig:approx} and Table~\ref{tab:approx}). It also retains the cancellation (median $\kappa_{\rm within}=5.38$, dominance in 192/200 points), because it keeps coherences above $T_\tau$ and in the $e$--$\mu$ sector. In contrast, keeping three resolved flavours or full coherence at all temperatures is inaccurate. RP-$M_1$ is slightly worse than RP-$T$.

Halving both thresholds changes the RP-$T$ asymmetry by a median 0.9\% and improves agreement near $T_\tau$ (Table~\ref{tab:criterion}).

We next turn to the scalar equations, for which SB-$T$ and SB-$M_1$ have median total errors of 55.4\%. The full $R_2+R_3$ contribution has median magnitude 52\% of $|\eta_B|$ and exceeds 20\% in 72\% of the points. SB-$T$ suppresses this contribution: its median heavier-parent error is 0.49 in units of $|\eta_B|$, versus 0.16 for $N_1$. In SB-$T$, the median relative $N_1$ error is 23\% for the 139 points below $M_1=10^{11}\,{\rm GeV}$ and 94\% above, where coherence matters near $T_\tau$~\cite{antusch2012}. Spectator variations are less important (Appendix~\ref{sec:rates}).

\begin{figure}[t]
\includegraphics[width=\columnwidth]{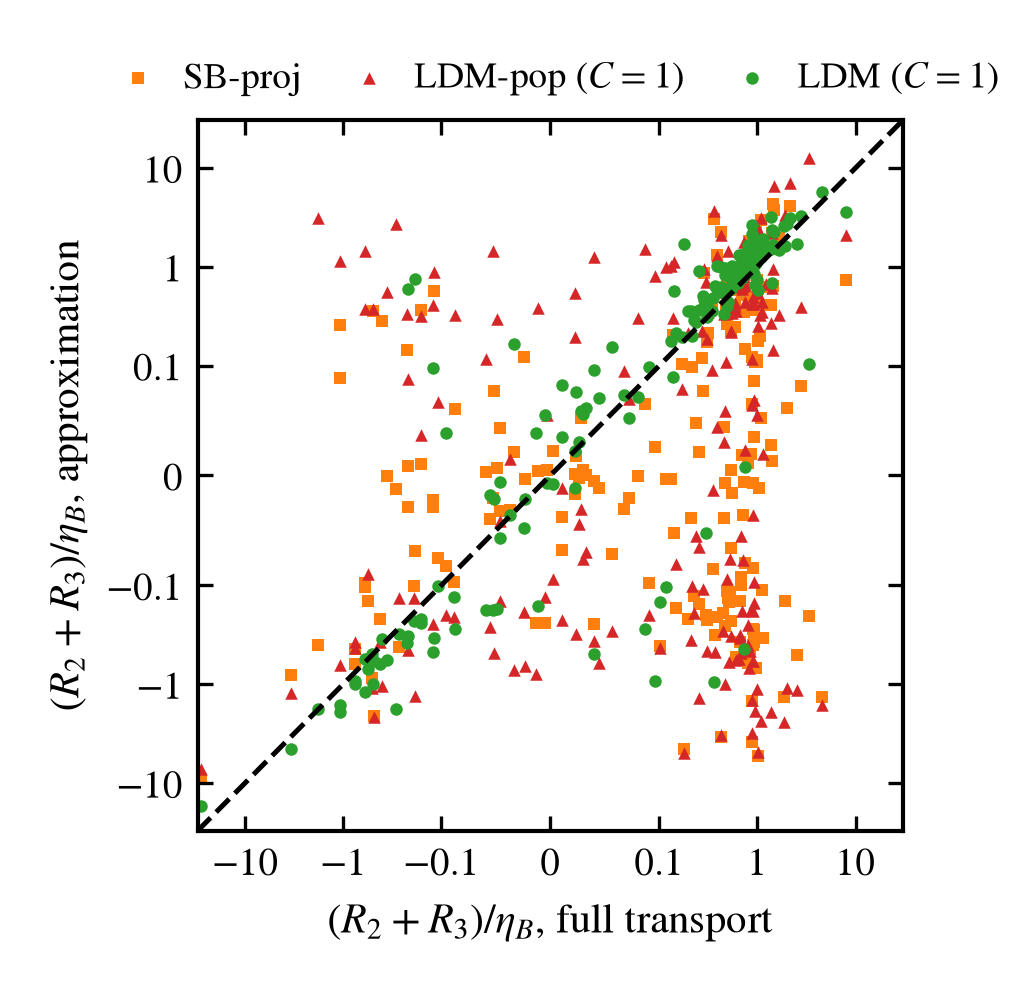}
\caption{Combined $N_2$ and $N_3$ contributions in the full and approximate evolutions, both normalized to the same full-transport $\eta_B^{\rm full}$ (linear scale within $\pm0.1$; dashed line: equality). LDM with full CP sources follows the full calculation most closely.}
\label{fig:heavier}
\end{figure}

In the 145 points where the full heavier-parent contribution exceeds 20\% of $|\eta_B|$, SB-proj and LDM-pop reproduce it with the correct sign and within a factor of two in only 23\% and 28\% of cases, compared with 87\% for LDM (Fig.~\ref{fig:heavier} and Table~\ref{tab:heavier}). LDM has a median total error of 16.7\%, compared with 5.2\% for RP-$T$, which retains dynamical spectators and the other charge-transport terms.

Above $T_\tau$, the LDM-pop source equals $\varepsilon^{(i)}_\parallel$ and drops $\varepsilon^{(i)}_\perp$. For an $N_2$ source followed by $N_1$ washout alone, Eq.~(\ref{eq:toy-response}) with $\overline P=\overline P_1$ shows how a trace-free source can acquire a net response. This is the contribution measured by $R_2^\perp$ in Eq.~(\ref{eq:invariant}) with all transport terms included. The paired LDM and LDM-pop comparison isolates the effect of replacing the full CP sources in the same evolution operator. Thus retaining a density matrix alone is not sufficient: the full CP source is crucial.

The partition-independent response of Sec.~\ref{sec:perp} provides a scale against which to compare the heavier-parent errors. In the 145 points where the full heavier-parent contribution exceeds 20\% of $|\eta_B|$, the median ratio of the error in $R_2+R_3$ to $|R_2^\perp+R_3^\perp|$ is 1.03 for SB-$T$, 1.08 for SB-proj, 1.17 for LDM-pop, and 0.47 for LDM. Unlike the paired LDM comparison, the SB errors also include differences in transport and threshold matching.

\begin{table}[!t]
\caption{Contribution of $N_2$ and $N_3$ in the 145 points where it exceeds 20\% of $|\eta_B|$ in the full transport: fractions of points in which an approximation has its sign and agrees with it within a factor of two. The density matrix with the full CP source has the highest agreement among the three approximations shown.}
\label{tab:heavier}
\begin{ruledtabular}
\begin{tabular}{lcc}
Equations & Same sign & Within factor two\\
\hline
SB-proj & 60\% & 23\%\\
LDM-pop, $C=1$ & 61\% & 28\%\\
LDM, $C=1$ & 97\% & 87\%
\end{tabular}
\end{ruledtabular}
\end{table}

Finally, even when an approximation reproduces the individual responses well, a large cancellation factor can magnify the error in their sum:
\begin{equation}
\frac{|\Delta\eta_B|}{|\eta_B|}=\kappa_{18}\frac{|\Delta\eta_B|}{\Lambda}.
\label{eq:amplification}
\end{equation}
For RP-$T$, the median shift in units of $\Lambda$ is only 0.72\%. Comparing the median values of $|\Delta\eta_B|/\Lambda$ and $|\Delta\eta_B|/|\eta_B|$, respectively, gives the following results for the full-transport cancellation factor: $\kappa_{18}<10$ (115 points: 0.71\% and 3.0\%) and $\kappa_{18}\geq10$ (85 points: 0.74\% and 15.7\%). The shift in units of $\Lambda$ is similar in the two groups, whereas the relative error is much larger in the second group (Fig.~\ref{fig:approx}b), illustrating Eq.~(\ref{eq:amplification}). The paired rate tests provide another illustration (Appendix~\ref{sec:rates}).

The RP-$T$ error grows with its own cancellation factor $\kappa_{18}^{\rm RP}$ and near $T_\tau$ (Table~\ref{tab:criterion}). For $\kappa_{18}^{\rm RP}\geq10$, 17.6\% of points are off by more than a factor of two. For $M_1<10^{11}\,{\rm GeV}$ and $\kappa_{18}^{\rm RP}<10$, 94.5\% of the points agree within 20\%. Outside this range the full calculation is preferable. This empirical criterion applies to RP-$T$ and the present sample only. For SB-$T$, the omission of much of the heavier-parent contribution leads to a large error in the responses themselves, as discussed above.

\begin{table}[!t]
\caption{Accuracy of RP-$T$ by mass and cancellation factor, where $\kappa_{18}^{\rm RP}$ is evaluated using RP-$T$ responses, with the median error for halved thresholds in the last column. Agreement worsens near the $\tau$ threshold and at large cancellation.}
\label{tab:criterion}
\begin{ruledtabular}
\footnotesize
\setlength{\tabcolsep}{2pt}
\begin{tabular}{lccccc}
$M_1$ (GeV) & $\kappa^{\rm RP}_{18}$ & Points & Median error & Within 20\% & Halved\\
\hline
$<10^{10}$ & all & 11 & 1.1\% & 100\% & 0.5\%\\
$10^{10}$--$10^{11}$ & all & 128 & 3.6\% & 86.7\% & 3.4\%\\
$10^{11}$--$10^{12}$ & all & 54 & 14.9\% & 64.8\% & 8.5\%\\
$>10^{12}$ & all & 7 & 134\% & 14.3\% & 55.7\%\\
\hline
$<10^{11}$ & $<10$ & 110 & 2.9\% & 94.5\% & --\\
all & $\geq10$ & 74 & 14.7\% & 62.2\% & --\\
\end{tabular}
\end{ruledtabular}
\end{table}

\section{Conclusions}\label{sec:conclusions}

We have shown that in the sampled thermal leptogenesis model with three hierarchical heavy neutrinos and decay CP sources, flavour-dependent washout leads predominantly to cancellations within individual parents rather than between them. The decomposition into 18 signed block responses establishes this dominance in the charged-lepton mass basis in both the calibrated and independent uncalibrated samples. The vanishing of the orthogonally projected CP matrix gives an analytic relation between the signs and sizes of the opposing diagonal and coherent responses in the coherent single-parent limit. Charged-lepton decoherence reduces their cancellation.

The trace-free response also has consequences for the heavier-neutrino contributions. The median signed ratio of the partition-independent trace-free response to the combined heavier-neutrino contribution is 84\% in the 145 points where that contribution exceeds 20\% of the final asymmetry in magnitude. The scalar population and overlap prescriptions tested largely miss this response, whereas a lepton density matrix with the full CP sources retains most of it. Thus the effect of flavour coherence is important not only for the survival of a pre-existing asymmetry or for a reversal of its sign, but also for the accuracy of the predicted baryon asymmetry in three-parent thermal leptogenesis.

\begin{acknowledgments}
The author used ChatGPT (OpenAI) and Claude (Anthropic) for assistance with numerical code development, figure preparation, and writing and editing of the manuscript. All calculations, results, and text were checked by the author, who takes full responsibility for the content of this paper.
\end{acknowledgments}

\appendix

\section{Calibration and sampling dependence}\label{sec:robustness}

For the calibration described in Sec.~\ref{sec:framework}, if the first choice of $\Omega$ does not yield a positive asymmetry, the CP branch $\Omega^*$ is tested, and a common mass-rescaling factor $f$ is then determined from the full transport to match the observed asymmetry $\eta_B^{\rm obs}$. At the proposal scale $f=1$, $M_1=M_1^{\rm prop}$. The search uses $10^{-8}\leq f\leq10^8$, a lower bound $M_1\geq10^7\,{\rm GeV}$, and $\sigma_{\max}(Y_\nu)\leq0.995$ after rescaling. A candidate is accepted only if $10^7\,{\rm GeV}\leq M_1\leq10^{16}\,{\rm GeV}$ and $\sigma_{\max}(Y_\nu)<1$. The scale is fixed by requiring that $\eta_B$ agree with $6.120\times10^{-10}$ within a relative tolerance of $2\times10^{-3}$. If this search does not find an acceptable $f$, the point is redrawn from the same priors; this happens for
28 points. 

Calibration does not use the cancellation factors. Table~\ref{tab:calibration} compares the calibration strata, the original scale, the opposite CP branch, and 200 independent points drawn without mass rescaling, branch selection or conditioning on $\eta_B$ (94 with positive $\eta_B$). Dominance in the latter sample is 182/200 (95\% binomial interval 86.1\%--94.6\%) and is largely lost without washout.

\begin{table}[t]
\caption{Calibration and uncalibrated samples; factors are medians and $N_{\rm dom}$ counts points with $\kappa_{\rm within}>\kappa_{\rm between}$ out of $N$ points in each ensemble or stratum. Within-parent dominance persists without calibration.}
\label{tab:calibration}
\begin{ruledtabular}
\scriptsize
\setlength{\tabcolsep}{2pt}
\begin{tabular}{lccc}
Ensemble or control & $\kappa_{\rm within}$ & $\kappa_{\rm between}$ & $N_{\rm dom}/N$\\
\hline
Calibrated, all & 6.66 & 1.06 & 196/200\\
\quad $f\leq1$ and $f>1$ & 5.67, 8.16 & 1.06, 1.06 & 106/109, 90/91\\
\quad $\Omega$ and $\Omega^*$ branch & 7.49, 5.93 & 1.12, 1.03 & 100/102, 96/98\\
\quad first draw and redrawn & 6.66, 6.88 & 1.06, 1.14 & 168/172, 28/28\\
Proposal scale $f=1$ & 7.21 & 1.04 & 195/200\\
Opposite CP branch & 6.42 & 1.11 & 192/200\\
\hline
Uncalibrated, all & 6.53 & 1.10 & 182/200\\
Uncalibrated, no washout & 1.27 & 1.81 & 48/200
\end{tabular}
\end{ruledtabular}
\end{table}

Reweighting the proposal mass from a log-uniform to a uniform prior (weights $w_k$ proportional to the proposal mass $M_1^{\rm prop}$ of point $k$, effective sample size $(\sum_kw_k)^2/\sum_kw_k^2=39.1$) changes the median $\kappa_{\rm within}$ from 6.66 to 7.92 and the dominance share from 98.0\% to 98.9\%.

We also examine whether the within-parent ordering could arise merely because six responses are summed before three. Keeping the 18 magnitudes of each point and assigning independent random signs, the probability of $\kappa_{\rm within}>\kappa_{\rm between}$ averages 0.737 (10th--90th percentiles 0.628--0.878), against the observed fraction 0.98. Drawing the signs instead from a Gaussian model with the $18\times18$ sign correlations of the other 199 points gives 0.801 (0.675--0.947).

\section{Reaction rates and basis dependence}\label{sec:rates}

To test the stability of within-parent dominance under changes in the transport, we perform paired calculations with a thermal initial abundance, with the top-Yukawa $\Delta L=1$ kernels scaled by 1.5 and 2, and with the $\Delta L=2$ coefficient scaled by 1.5. Dominance persists in 195--196 of the 200 points (Table~\ref{tab:sensitivity}). A thermal initial abundance has the largest effect on $\eta_B$, with a 90th-percentile change of 76.3\% and a sign reversal in 6 points.

\begin{table}[t]
\caption{Paired sensitivity tests for the 200 points: thermal instead of zero initial abundance, top-Yukawa $\Delta L=1$ kernels scaled by 1.5 and 2, and $\Delta L=2$ coefficient scaled by 1.5. $\Delta_{\rm ch}=\sum_{i,a}|\Delta R_{i,a}|/\Lambda$ is the change of the response pattern; all entries except $N_{\rm dom}$ (the number of points with $\kappa_{\rm within}>\kappa_{\rm between}$) are medians. Dominance survives every variation, and the net asymmetry is most sensitive to the initial abundance.}
\label{tab:sensitivity}
\begin{ruledtabular}
\footnotesize
\setlength{\tabcolsep}{2pt}
\begin{tabular}{lccccc}
Test & $\Delta_{\rm ch}$ & $|\Delta\eta_B/\eta_B|$ & $\kappa_{\rm within}$ & $\kappa_{\rm between}$ & $N_{\rm dom}$\\
\hline
Baseline & 0 & 0 & 6.66 & 1.06 & 196\\
Thermal $N_i$ & 0.355 & 13.8\% & 7.93 & 1.17 & 195\\
$\Delta L=1$, $\times1.5$ & 0.032 & 1.3\% & 6.79 & 1.07 & 196\\
$\Delta L=1$, $\times2$ & 0.056 & 2.4\% & 7.00 & 1.08 & 196\\
$\Delta L=2$, $\times1.5$ & $1.0\times10^{-5}$ & $2.9\times10^{-3}$\% & 6.66 & 1.06 & 196\\
\end{tabular}
\end{ruledtabular}
\end{table}

For example, doubling the top-Yukawa $\Delta L=1$ kernels gives median values of $|\Delta\eta_B|/\Lambda$ and $|\Delta\eta_B|/|\eta_B|$ of 0.33\% and 2.4\%, respectively, using the baseline $\Lambda$ and $\eta_B$. The larger relative change reflects the amplification in Eq.~(\ref{eq:amplification}). In RP-$T$, multiplying the quark-Yukawa and sphaleron rates by ten changes the $N_1$ contribution by a median 6.9\%; replacing $C$ by the unit matrix in the scalar equations likewise has little effect on the median errors (Table~\ref{tab:approx}).

\begin{table*}[!t]
\caption{Cancellation factors for different partitions. Dominance counts points with a factor greater than $\kappa_{\rm between}$; Null gives the corresponding probability for independent random signs. The washout-aligned partition has little cancellation.}
\label{tab:partitions}
\begin{ruledtabular}
\begin{tabular}{lccc}
Partition (full transport unless stated) & Median factor & Dominance & Null\\
\hline
Six charged-lepton blocks, $\kappa_{\rm within}$ & 6.66 & 196/200 & 0.737\\
Diagonal and coherent sums, $\kappa_{\rm d/o}$ & 4.89 & 193/200 & 0.633\\
Six blocks of the $f_1$ and $f_2$ pieces & 6.46 and 13.83 & -- & --\\
Washout-aligned split, Eq.~(\ref{eq:aligned-split}) & 1.29 & 129/200 & 0.514\\
Random flavour bases (15 points $\times$ 32 bases) & 31.33 & 480/480 & --\\
\hline
Three diagonal blocks, diagonal charges only & 1.09 & 129/200 & 0.679
\end{tabular}
\end{ruledtabular}
\end{table*}

As a numerical resolution check, we repeated the 18-channel transport at points $k=0,77,150$ using 200 and 400 instead of 100 logarithmically spaced interpolation nodes in $z$. The change in the sum of absolute channel differences, divided by the 100-node value of $\Lambda$, was below $6\times10^{-5}$ for these tests. At the same three fixed masses we also tested a factorized source ansatz: replacing $D_i(Y_{N_i}-Y_{N_i}^{\rm eq})$ in Eq.~(\ref{eq:source}) by $-dY_{N_i}/dz$ amounts to assigning the decay CP matrix to every production and destruction reaction in Eq.~(\ref{eq:heavy}). This ansatz is motivated by the approximate relation between decay and scattering CP asymmetries in hierarchical regimes~\cite{nardiracker2007}, but is not a calculation of the scattering CP matrices. At 400 nodes the respective changes in $\eta_B$ were $+12.8\%$, $+9.4\%$, and $-22.1\%$; within-parent cancellation remained dominant in all three points.

We also check the analytic relations and limiting cases used in the main text. The CP identity~(\ref{eq:identity}) holds to better than $10^{-14}$ throughout the ensemble, and the $f_2$ piece is traceless to better than $10^{-13}$. In the two-flavour model of Eq.~(\ref{eq:toy}), the damping profile is illustrative; the charged-lepton transfer coefficient in the full transport is approximately constant in $z$ for fixed couplings.
The coherent solutions agree with Eq.~(\ref{eq:toy-kappa}) to $1.8\times10^{-8}$, and Eq.~(\ref{eq:toy-fast}) agrees with the numerical solution to 5.8\% at $\Gamma_0=10$ and 0.06\% at $\Gamma_0=10^3$.
The equilibrium limit of the Standard Model operator $A$ with fast $\tau,c,b,t$ Yukawa interactions and both sphalerons, and absent $e,\mu,u,d,s$ Yukawa interactions, reproduces $C^{(2)}$ to better than $10^{-10}$. With all flavours resolved, the LDM and LDM-pop equations reduce numerically to the scalar equations with $C=1$ to better than $10^{-6}$.

To quantify the dependence on the basis, the CP matrices of 15 points ($k$ = 0, 5, 28, 42, 54, 81, 91, 115, 121, 126, 146, 159, 179, 186, 190), whose ranks in $\kappa_{\rm within}$ within each mass ratio are closest to the 5th, 25th, 50th, 75th, and 95th percentiles, are repartitioned in 32 random flavour bases each, distributed uniformly over the unitary group. The transport is fixed: only the partition of the CP source changes, while its total matrix, the physical charged-lepton basis, the operator $A(z)$, the heavy-neutrino abundances, and the conversion to $\eta_B$ are kept. Because the response of each parent is linear in its source matrix, it follows from the six block responses and three further off-diagonal source directions. The median $\kappa_{\rm within}$ of these points is 6.49 in the charged-lepton basis (95\% bootstrap interval from 10\,000 resamplings of the 15 points 3.03--12.54) and 31.33 over random bases (11.66--37.38), with $\kappa_{\rm within}>\kappa_{\rm between}$ in all 480 repartitions, while the parent totals change by less than $10^{-12}$.

Table~\ref{tab:partitions} compares the partitions; only the initial-abundance and rate variations recompute the heavy-neutrino histories.

\section{Numerical implementation}\label{sec:numerical}

For completeness, we give the numerical prescription for the rates and integration used in Sec.~\ref{sec:framework}. The Standard Model transfer rates retain the parametrizations of Ref.~\cite{bolehsoftware}. For the lepton-number-violating rates, we use the Standard Model thermal masses of Eqs.~(83a)--(83c) and (83e) of Ref.~\cite{giudice2004}, with the fixed couplings of Sec.~\ref{sec:framework} and the right-handed top thermal mass set equal to the quark-doublet thermal mass. The top-Yukawa and gauge scattering cross sections are Eqs.~(94)--(97) of that reference, including its $U(1)_Y$ contribution and average of the two gauge $t$ channels. Let $x$ be the squared centre-of-mass energy divided by $M_i^2$, $x_{\min}$ the larger of the initial and final two-body thresholds in these units, and $\sigma_{H_{s,t}}$ the reduced cross sections divided by $h_{ii}y_t^2$ (by $h_{ii}$ alone for the analogous gauge integrals). The integrals in Eq.~(\ref{eq:thermal-rates}) are
\begin{equation}
I_{s,t}(z_i)=z_i^3\int_{x_{\min}}^\infty dx\,\sqrt{x}\,\sigma_{H_{s,t}}(x){\rm K}_1(z_i\sqrt{x}).
\label{eq:thermal-integrals}
\end{equation}
Mass ratios are evaluated as specified in Sec.~\ref{sec:framework}. With $L$ the lepton doublet and $H$ the Higgs, the factor $F_{\rm QP}$ is obtained from the $LH\leftrightarrow\bar L\bar H$ and $LL\leftrightarrow HH$ collision integrals, including thermal masses and quantum statistics, normalized to $T^3{\rm Tr}(c_\nu c_\nu^\dagger)/\pi^3$.

\end{document}